\documentclass[prx,aps,showpacs,twocolumn,preprintnumbers,
amsmath,amssymb,superscriptaddress,longbibliography]{revtex4-1}
\usepackage[english]{babel}
\usepackage{amsmath,amssymb,amsfonts}
\usepackage{graphicx}
\usepackage[colorlinks=True,linkcolor=red,citecolor=blue,urlcolor=blue]{hyperref}
\usepackage{bookmark}
\usepackage[dvipsnames]{xcolor}
\usepackage{chngcntr}
\usepackage{braket}
\usepackage{nicefrac}
\usepackage{bm}
\usepackage{bbm}
\usepackage{braket}
\usepackage{slashed}
\usepackage[normalem]{ulem}
\usepackage{array}
\newcolumntype{C}{>{$}c<{$}}
\usepackage{chemmacros}

\definecolor{Blue}{rgb}{0.,0.24,0.51}
\newcommand{\Blue}{\color{Blue}}

\definecolor{Nathanpink}{rgb}{0.94,0.317,0.9607}

\begin{document}

%\title{Orbital and heat magnetizations: Generalized sum rules and dichroic measurements}

%\title{{\Blue Dichroism from Thermoelectric Chiral Drives: \\ Generalized Sum Rules for Orbital and Heat Magnetizations}}

\title{{\Blue Dichroism from Chiral Thermoelectric Probes: \\ Generalized Sum Rules for Orbital and Heat Magnetizations}}

\author{Baptiste Bermond}
\email{baptiste.bermond@lkb.ens.fr}
\affiliation{Laboratoire Kastler Brossel, Coll\`ege de France, CNRS, ENS-Universit\'e PSL,
Sorbonne Universit\'e, 11 Place Marcelin Berthelot, 75005 Paris, France
}
\author{Lucila {Peralta Gavensky}}
\email{lucila.peralta.gavensky@ulb.be}
\affiliation{International Solvay Institutes, 1050 Brussels, Belgium}
\affiliation{Center for Nonlinear Phenomena and Complex Systems, Universit\'e Libre de Bruxelles, CP 231, Campus Plaine, B-1050 Brussels, Belgium}
\author{Ana\"is Defossez}
\email{anais.defossez@lkb.ens.fr}
\affiliation{Laboratoire Kastler Brossel, Coll\`ege de France, CNRS, ENS-Universit\'e PSL,
Sorbonne Universit\'e, 11 Place Marcelin Berthelot, 75005 Paris, France
}
\affiliation{International Solvay Institutes, 1050 Brussels, Belgium}
\affiliation{Center for Nonlinear Phenomena and Complex Systems, Universit\'e Libre de Bruxelles, CP 231, Campus Plaine, B-1050 Brussels, Belgium}
\author{Nathan Goldman}
\email{nathan.goldman@lkb.ens.fr}
\affiliation{Laboratoire Kastler Brossel, Coll\`ege de France, CNRS, ENS-Universit\'e PSL,
Sorbonne Universit\'e, 11 Place Marcelin Berthelot, 75005 Paris, France
}
\affiliation{International Solvay Institutes, 1050 Brussels, Belgium}
\affiliation{Center for Nonlinear Phenomena and Complex Systems, Universit\'e Libre de Bruxelles, CP 231, Campus Plaine, B-1050 Brussels, Belgium}

%%%%%%%%%%%%%%%%%%%%%%%%%%%%%%%%%%%%%%%%%%%%%%%%%%%%%%%%%%%%%%%%%%%%%%%%%%%%%
\begin{abstract}
We introduce a unified framework that relates orbital and heat magnetizations to experimentally accessible excitation spectra, through thermoelectric probes and generalized sum rules. By analyzing zero-temperature transport coefficients and applying Kramers-Kronig relations, we derive spectral representations of magnetization densities from thermoelectric correlation functions. Excitation rates under chiral thermoelectric drives then naturally emerge as direct probes of these Kubo-type correlators, placing orbital and heat magnetizations on equal footing with the topological  Chern number. As a direct consequence of our formalism, we introduce a hierarchical construction that organizes orbital and heat magnetizations into distinct physical contributions accessible through sum rules, and also naturally obtain real-space markers of these magnetizations. Besides, non-chiral thermal probes identify a heat quantum metric, which is defined over the space of gravitomagnetic deformations. From an experimental standpoint, we propose concrete implementations of thermoelectric dichroic measurements in quantum-engineered platforms based on modulated strain fields. These results establish thermoelectric dichroic measurements as a versatile route to access and disentangle fundamental ground-state properties.
\end{abstract}
\date{\today}
\maketitle

\section{Introduction}

%\subsection{General introduction}

Sum rules establish fundamental connections between excitation spectra and ground state properties, making them essential tools for understanding and interpreting the physics of condensed matter systems~\cite{kubo1957statistical,bennett1965faraday,kubo1972kramers,mazenko2006nonequilibrium,sakurai2020modern,pitaevskii2016bose}. In recent years, sum rules have garnered renewed attention for their ability to probe geometric and topological properties in many-body quantum systems~\cite{yu2024quantum}. Indeed, the quantum metric and the Berry curvature -- as well as their associated topological invariants -- can be expressed in terms of sum rules involving dissipative responses upon driving~\cite{souza2008dichroic,tran2017probing,ozawa2018extracting,ozawa2019probing,repellin2019detecting,schuler2020local,goldman2024relating,verma2024instantaneous,bermond2025local}. This suggests practical methods to probe these geometric properties in a wide range of physical settings, including solid-state devices~\cite{yu2020experimental,tan2019experimental,klees2020microwave,chen2022synthetic,beaulieu2024berry,kang2025measurements} and ultracold gases~\cite{asteria2019measuring}. Besides, sum rules have also emerged as fundamental identities for establishing bounds on material properties~\cite{kruchkov2023spectral,onishi2024topological,onishi2024fundamental,verma2024instantaneous,kruchkov2025bounds,shinada2025quantum,ji2025density}. 

Despite their utility in capturing various aspects of quantum matter~\cite{yu2024quantum}, these sum rules mostly involve the electric conductivity tensor and its integrals over frequencies, $\int_0^{+\infty}\omega^n\sigma^{\mu\nu}(\omega) d\omega$. As a result, this framework has been essentially confined to properties related to electric transport, or, from the perspective of Kubo's linear response theory~\cite{kubo1957statistical}, to electric current-current correlation functions $\langle\hat{J}^\mu_e\hat{J}^\nu_e\rangle $. A natural extension would be to explore a broader class of sum rules that encompass transport phenomena of a different nature, such as thermal transport~\cite{Cooper1997,qin2011energy,Auerbach2024}. This direction is particularly compelling in systems hosting neutral collective excitations, such as spin waves (magnons) or propagating Majorana  modes in spin systems~\cite{katsura2010theory,matsumoto2011rotational,matsumoto2011theoretical,nasu2017thermal}. In this context, the relevant correlation functions become $\langle \hat{J}_e^\mu\hat{J}_Q^\nu \rangle $ (electric-heat current correlations) and $\langle\hat{J}_Q^\mu\hat{J}_Q^\nu\rangle$ (heat-heat current correlations), where $\hat{J}_Q$ denotes the heat current. Identifying which physical quantities are encoded in sum rules built from thermoelectric correlations remains an intriguing challenge, offering rich opportunities from both fundamental and applied perspectives.

Exploring generalized sum rules associated with thermal transport and probes naturally brings us to the notions of orbital and heat magnetizations. These fundamental ground-state properties are central to connecting thermal transport coefficients with Kubo-type   correlation functions~\cite{Cooper1997,qin2011energy,Auerbach2024}. In Bloch-band systems, such magnetizations admit a clear interpretation in terms of electric or heat magnetic moments -- arising from the self-rotation of electronic wave packets -- and circulating edge currents~\cite{xiao2005berry,thonhauser2005orbital,ceresoli2006orbital,THONHAUSER_2011,qin2011energy}. Specific physical contributions to these magnetizations can be formulated as sum rules involving the electric conductivity tensor~\cite{kubo1957statistical,souza2008dichroic}. For example, the self-rotation component of orbital magnetization can be obtained via the dichroic 
$f$-sum rule~\cite{oppeneer1998magneto,verma2024instantaneous,souza2008dichroic,ji2025density}. Yet, how the total orbital and heat magnetizations emerge from sum rules, and how their individual contributions can be systematically accessed through thermoelectric dichroic measurements, has been largely overlooked. Addressing this gap constitutes a primary motivation for this work.

In this work, we establish a unified framework that connects ground-state topological and magnetization properties to experimentally accessible excitation spectra. By analyzing the zero-temperature limit of transport coefficients, and making use of the Kramers–Kronig relations, we demonstrate that orbital and heat magnetization densities admit spectral representations derived from thermoelectric and thermal correlation functions. Building on this insight, we propose that frequency-integrated excitation rates under suitably engineered chiral drives serve as powerful probes for extracting Kubo-type coefficients, thereby enabling experimental access to the full set of ground-state properties. In particular, while purely electrical drives yield the topological Chern number  of an insulating state (i.e.~its quantized Hall response), mixed and purely thermal drives reveal the orbital and heat magnetization densities, placing these quantities on equal footing within integrated dichroic spectra. Moreover, we discuss how physically distinct contributions to the orbital and heat magnetizations can be identified by combining a proper set of sum rules. We present a generalization of this partitioning through a hierarchical construction, which organizes generalized (higher-order) magnetizations according to sum rules. We illustrate our results on realistic models, and describe how this thermoelectric dichroic approach can be implemented in quantum engineered settings, such as cold atoms in optical lattices.

\subsection*{Outline}

The rest of the paper is organized as follows:~To set the stage, Section~\ref{section_illustration} illustrates the connection between sum rules for transport coefficients and dichroic measurements. Section~\ref{section_sum_rules} then introduces sum rules for thermoelectric transport, deriving explicit relations for the orbital and heat magnetizations in terms of Kubo correlation functions, and discussing their connection to the dichroic $f$-sum rule. We derive additional sum rules that characterize the ground-state quantum fluctuations of electric and thermal polarizations, leading to the notion of a ``heat quantum metric".
In Section~\ref{Section_generalized_dichroism}, we explain how these sum rules can be probed through thermoelectric dichroic measurements, i.e.~by monitoring excitation rates upon driving the system with suitably designed thermoelectric probes (chiral and non-chiral). Section~\ref{section_open_exp} presents a real-space formulation that naturally yields local markers for orbital and heat magnetizations, and explores experimental strategies for implementing thermoelectric dichroic measurements. In particular, we propose a protocol to resolve individual contributions to the total orbital magnetization. Finally, Section~\ref{section_generalization} shows how sum rules enable a natural partitioning of the heat magnetizations, leading to a hierarchical construction of generalized (higher-order) heat magnetizations and their corresponding sum rules. Perspectives and outlook are discussed in Section~\ref{section_conclusion}.

\section{Transport coefficients, sum rules and dichroic measurements: a brief illustration}\label{section_illustration}

We first motivate our approach by illustrating the interconnections between sum rules for transport coefficients and dichroic measurements, which form the foundation of the generalized sum rules investigated in this work; see also Refs.~\cite{bennett1965faraday,souza2008dichroic,tran2017probing,goldman2024relating}.

Let us consider a two-dimensional insulator subjected to a circularly polarized electric field, $\bm{E}_\pm\!=\!2E_0 (\cos(\omega t),\pm\sin(\omega t),0)$. In the dipolar approximation, one can include this perturbation in the Hamiltonian through a vector potential $\bm{A}_\pm\!=\!(2E_0/\omega) (\sin(\omega t),\mp\cos(\omega t),0)$ minimally coupled to the electric current operator,
\begin{align}
    \delta\hat{\mathcal{H}}^\pm
    &=\frac{2E_0}{\omega}\left[\sin(\omega t)\hat{J}^x_e\mp\cos(\omega t)\hat{J}^y_e\right],\label{eq:perturbation}
\end{align}
where $\hat{J}_e^{\mu}\!=\!-e \hat v^{\mu}$ denotes the components of the single-particle electric current operator. Following Bennett and Stern~\cite{bennett1965faraday}, the power absorbed $\mathcal{P}^\pm(\omega)$ by an isotropic system of area $A$ subjected to the circularly polarized electric field $\bm{E}_\pm$ can be expressed in terms of the absorptive part of the optical conductivity tensor, $\sigma^{{\rm abs}}(\omega)=(\sigma(\omega)+\sigma^{\dagger}(\omega))/2$, according to
\begin{equation}
    \mathcal{P}^\pm(\omega)\!=\!4E_0^2 A\left ({\rm Re}\left[\sigma^{{\rm abs}, xx}(\omega)\right ]    \mp   {\rm Im}\left[\sigma^{{\rm abs},xy}(\omega)\right]\right ),
\end{equation}
where ${\rm Re}/{\rm Im}$ denote real/imaginary parts. Hence, the absorptive part of the conductivity tensor can be related to the differential power absorbed, $\Delta \mathcal{P}(\omega)\!=\!(\mathcal{P}^+(\omega)-\mathcal{P}^-(\omega))/2$, through the  useful relation 
\begin{align}
\label{eq:power_absorption}
    {\rm Im}\left [\sigma^{{\rm abs},xy}(\omega)\right]&= - \frac{\Delta \mathcal{P}(\omega)}{4 A E_0^2 } = - \frac{\hbar \omega \Delta \Gamma(\omega) }{4 A E_0^2}.
\end{align}
This dissipative part of the conductivity tensor thus reflects the circular dichroism of the medium. For later purposes, we express the power absorbed in terms of an excitation rate, 
\begin{equation}
\Gamma^{\pm} (\omega) = \mathcal{P}^{\pm} (\omega) /\hbar \omega ,
\end{equation}
which reflects the rate at which a photon $\hbar \omega$ is absorbed by the system, and we introduced the corresponding differential rate $\Delta \Gamma\!=\!(\Gamma^+ - \Gamma^-)/2$ in Eq.~\eqref{eq:power_absorption}.

We now apply Kramers-Kronig relations~\cite{bennett1965faraday} to the imaginary part of the conductivity tensor in Eq.~\eqref{eq:power_absorption}, 
\begin{align}
\sigma^{xy}&=\underset{\omega\to0}{\lim}{\rm Re}[\sigma^{xy}(\omega)]=\frac{2}{\pi}\int_0^{\infty}\frac{{\rm Im} [\sigma^{{\rm abs},xy}(\omega)]}{\omega} d \omega \notag \\
&= - \frac{\hbar}{2 \pi E_0^2A} \Delta \Gamma^{{\rm int}}\,,\label{eq:sum_rule_sigma}
\end{align}
where we introduced the integrated differential rate  
\begin{equation}
\Delta \Gamma^{{\rm int}}\!\equiv\!(1/2) \int_0^{\infty} \left [\Gamma^{+}(\omega) - \Gamma^{-}(\omega) \right ] d\omega .
\end{equation}
The sum rule in Eq.~\eqref{eq:sum_rule_sigma} provides an explicit relation between the DC Hall conductivity $\sigma^{xy}$ and the circular dichroism of the medium, captured by the differential rates $\Delta \Gamma\!=\!(\Gamma^+ - \Gamma^-)/2$. 

For a two-dimensional Chern insulator, the Hall conductivity is quantized according to the many-body Chern number~\cite{niu1985quantized}, such that the integrated differential rate $\Delta \Gamma^{{\rm int}}$ in Eq.~\eqref{eq:sum_rule_sigma} is quantized~\cite{tran2017probing,repellin2019detecting,goldman2024relating}. This quantized dissipative effect was first observed in an ultracold atomic gas~\cite{asteria2019measuring}. For a wave packet prepared in a Bloch band, this circular-dichroic approach gives access to the local Berry curvature of the band~\cite{tran2017probing,schuler2020local}, as was experimentally demonstrated in Refs.~\cite{beaulieu2024berry,bac2025probing}.

%%%%%%%%%
%%%%%%%%%

The excitation rates $\Gamma^{\pm}(\omega)$ can be explicitly evaluated within the framework of Fermi's golden rule~\cite{tran2017probing,goldman2024relating}, considering the time-dependent perturbation in Eq.~\eqref{eq:perturbation}
\begin{align}
&\Gamma^{\pm} (\omega)\!=\! \frac{2 \pi}{\hbar}\sum_{n\in \textrm{occ}} \, \sum_{m\in \textrm{unocc}} \, \vert \mathcal{V}_{mn}^\pm\vert ^2 \, \delta(\varepsilon_m \!-\! \varepsilon_n  \!-\!  \hbar\omega), \label{eq_FGR} \\
&\vert \mathcal{V}_{mn}^\pm \vert ^2=\left (E_0/\hbar \omega \right )^2 \bigg \vert \bigg \langle m \bigg \vert  \hat{J}^x_e\!\pm\! i\hat{J}^y_e \bigg \vert n   \bigg \rangle \bigg \vert^2 , \notag
\end{align}
where $\{ \vert n \rangle \}$ (resp.~$\{ \vert m \rangle \}$) denotes the set of occupied (resp.~empty) single-particle eigenstates of energy $\varepsilon_n$ (resp.~$\varepsilon_m$). By substituting the expression~\eqref{eq_FGR} into the sum rule given in Eq.~\eqref{eq:sum_rule_sigma}, one recovers Kubo's formula for the Hall conductivity, expressed as a current-current correlation function~\cite{kubo1957statistical}:
\begin{equation}
\sigma^{xy}=\frac{1}{iA} \sum_{n\in \textrm{occ}} \, \sum_{m\in \textrm{unocc}} \frac{\langle n \vert \hat J^x_e \vert m \rangle \langle m \vert \hat J^y_e \vert n \rangle - (x \leftrightarrow y)}{(\varepsilon_m \!-\! \varepsilon_n)^2} .
\end{equation}
This example illustrates how Kubo-type responses, linked to ground-state transport properties, can be connected to a dissipative dichroic response and experimentally identified by monitoring the excitation rates $\Gamma_{\pm} (\omega)$ -- or power absorbed $\mathcal{P}^{\pm} (\omega)$ -- under external driving.

We close this introductory section by remarking that the circular perturbation in Eq.~\eqref{eq:perturbation} can be equally introduced through a direct coupling to the electric polarization, 
\begin{align}
    \delta\hat{\mathcal{H}}^\pm
    &=2E_0\left[\cos(\omega t)\hat{P}^{x}_e\pm\sin(\omega t)\hat{P}^{x}_e\right],\label{eq:perturbation_one}
\end{align}
where $\hat{P}^{\mu}_e\!=\!-e\,\hat{r}^{\mu}$ denotes the components of the single-particle electric polarization operator. This circular drive can be naturally realized in neutral atomic gases by periodically shaking the lattice~\cite{struck2011quantum,jotzu2014experimental}. In these engineered settings, the resulting excitation rates can be extracted from band-mapping images, which reveal the redistribution of atoms among the Bloch bands under driving~\cite{asteria2019measuring}.

\section{Thermoelectric transport coefficients and sum rules}\label{section_sum_rules}

\subsection{General framework}

Linear response theory provides a systematic way to express electrical and thermal transport coefficients in terms of equilibrium correlation functions. Within this framework, the DC electric ($\bm{J}_e$) and heat ($\bm{J}_Q$) currents are related to applied static electric fields $\bm{E}$ and temperature gradients $-\bm{\nabla}T$ via

\begin{eqnarray}
    \begin{pmatrix}\bm{J}_e\\
    \bm{J}_Q\end{pmatrix}
    &=&
    \begin{pmatrix}
        \sigma & \alpha_{\mathrm{th}}\\
        T\,\overline{\alpha}_{\mathrm{th}} & \kappa
    \end{pmatrix}
    \begin{pmatrix}
        \bm{E}\\
        -\bm{\nabla}T
       \end{pmatrix}.
       \end{eqnarray}
Here $\sigma$ denotes the electrical conductivity tensor,  
$\alpha_{\mathrm{th}}$ and $\overline{\alpha}_{\mathrm{th}}$ the thermoelectric tensors leading to Seebeck and Peltier effects,  and $\kappa$ the thermal conductivity tensor.  Following Refs.~\cite{Cooper1997,qin2011energy,Auerbach2024}, these transport coefficients can be obtained from the zero-frequency limits of the corresponding Kubo correlation functions as:
\begin{eqnarray}
\notag
    \sigma &=& \mathbb{L}_{ee}^{\mathrm{DC}}\\
    \notag
    \alpha_{\mathrm{th}} &=& \frac{1}{T}\left(\mathbb{L}_{eQ}^{\mathrm{DC}} - c\,\bm{\mathcal{M}} \times\right)\\
    \notag
    \overline{\alpha}_{\mathrm{th}} &=& \frac{1}{T}\left(\mathbb{L}_{Qe}^{\mathrm{DC}} - c\,\bm{\mathcal{M}} \times\right)\\
    \label{transp_coeff}
    \kappa &=& \frac{1}{T}\left(\mathbb{L}_{QQ}^{\mathrm{DC}} - 2\,\bm{\mathcal{M}}^{Q} \times\right),
\end{eqnarray}
where  $\bm{\mathcal{M}}$ and $\bm{\mathcal{M}}^{Q}$ denote the orbital and heat magnetization densities of the system. In these expressions, $\bm{\mathcal{M}}\times$ represents the linear operator associated with the cross product by $\bm{\mathcal{M}}$; in components  $(\mathcal{M}\times)_{\mu\nu}=\epsilon_{\mu\rho\nu}\mathcal{M}_{\rho}$, and similarly for $\bm{\mathcal{M}}^{Q}\times$.
In analogy with the orbital magnetization, which is associated with the moment $\bm{r}\times \bm{j}_e(\bm{r})$ of the charge current density, the heat magnetization can be defined through the moment $\bm{r}\times \bm{j}_Q(\bm{r})$ of the heat current density~\cite{zhang2020thermodynamics}.

The DC limit of the Kubo matrices is defined as 
\begin{equation}
   \mathbb{L}_{\alpha\beta}^{\mathrm{DC}}=\lim_{\omega+i\epsilon \to 0}\lim_{A\to \infty} \mathrm{Re}[\mathbb{L}_{\alpha\beta}(\omega)] ,
   \label{Lmunu_DC}
\end{equation}
where the finite-frequency correlation functions are defined as~\cite{Auerbach2024}
\begin{equation}
    \mathbb{L}_{\alpha\beta}^{\mu\nu}(\omega)
    = -\frac{i}{\hbar A}\sum_{nm}
    \frac{\langle n|\hat{J}_{\alpha}^{\mu}|m\rangle
    \langle m|\hat{J}_{\beta}^{\nu}|n\rangle}
    {\omega+i\epsilon+(\omega_n -\omega_m)}
    \frac{f_n -f_m}{\omega_n-\omega_m}.
    \label{Lmunu}
\end{equation}
For simplicity, we restrict the discussion to non-interacting fermionic systems, where $|n\rangle$ are single-particle eigenstates of the Hamiltonian $\hat{H}$ with energies $\varepsilon_n = \hbar \omega_n$ and $f_n$ denotes the Fermi–Dirac occupation. Spatial indices $\mu,\nu\in{x,y}$ label Cartesian components of the current operators $\hat{\bm{J}}_{\alpha}$, while the indices
\begin{equation}
\alpha,\beta = e,Q ,
\end{equation}
distinguish between electric and heat currents. We note that the electric-electric response tensor coincides with the optical conductivity introduced in Sec.~\ref{section_illustration}, namely $\mathbb{L}^{\mu\nu}_{ee}(\omega)\equiv \sigma^{\mu\nu}(\omega)$.

The subtractions involving orbital and heat magnetizations in Eq.~\eqref{transp_coeff} ensure that the transport coefficients describe the response of true transport currents to external fields, excluding contributions from circulating magnetizations~\cite{qin2011energy,Auerbach2024,Smrcka1977,Cooper1997}. As we will show in the following, the Kubo approach provides a natural starting point for deriving exact frequency-integrated constraints, or \emph{sum rules}, which follow from causality and the Kramers–Kronig relations.

The current operators $\hat{\bm{J}}_{\alpha}$ can be expressed as time derivatives of the corresponding polarization operators $\hat{\bm{P}}_{\alpha}$,
\begin{equation}
\hat{J}_{\alpha}^{\mu} = \frac{i}{\hbar}[\hat{H},\hat{P}_{\alpha}^{\mu}],
\end{equation}
which provides a convenient framework for formulating response functions in terms of polarization matrix elements. Under open boundary conditions, the one-body polarization operators are well-defined and take the real-space forms
\begin{subequations}
\label{pol_real}
    \begin{align}
        \hat{P}_e^{\mu} &= -e\,\hat{r}^{\mu},\\
        \hat{P}_Q^{\mu} &= \tfrac{1}{2}\!\left\{\hat{r}^{\mu},\hat{H}-\mu\right\},
    \end{align}
\end{subequations}
corresponding to electric and heat polarizations, respectively.

We will focus on insulating states, for which only interband matrix elements ($n \neq m$) contribute to the Kubo response in Eq.~\eqref{Lmunu}.
Using the relation $\langle n|\hat{J}_{\alpha}^{\mu}|m\rangle = i\omega_{nm}\langle n|\hat{P}_{\alpha}^{\mu}|m\rangle$, the frequency-resolved correlation functions can then be written as
\begin{equation}
\mathbb{L}_{\alpha\beta}^{\mu\nu}(\omega) = \frac{i}{\hbar A}\sum_{n\neq m} f_{nm}\omega_{mn}\frac{\langle n|\hat{P}^{\mu}_{\alpha}|m\rangle \langle m|\hat{P}^{\nu}_{\beta}|n\rangle}{\omega - \omega_{mn} + i \epsilon},
\label{inter_Lee}
\end{equation}
where $\omega_{nm} = \omega_{n} - \omega_m$ and $f_{nm} = f_{n} - f_m$. The product of matrix elements of the polarization operators in Eq.~\eqref{inter_Lee} can be further decomposed into symmetric and antisymmetric parts as
\begin{equation}
   \langle n|\hat{P}^{\mu}_{\alpha}|m\rangle \langle m|\hat{P}^{\nu}_{\beta}|n\rangle = P^{+,\mu\nu}_{\alpha\beta,nm}+i\frac{P_{\alpha\beta,nm}^{-,\mu\nu}}{2},
   \label{P_decomp}
\end{equation}
with
\begin{eqnarray}
\notag
    P^{+,\mu\nu}_{\alpha\beta,nm} &=& \mathrm{Re}\left[\langle n|\hat{P}^{\mu}_{\alpha}|m\rangle \langle m|\hat{P}^{\nu}_{\beta}|n\rangle\right]\\ 
    P^{-,\mu\nu}_{\alpha\beta,nm}&=& 2\mathrm{Im}\left[\langle n|\hat{P}^{\mu}_{\alpha}|m\rangle \langle m|\hat{P}^{\nu}_{\beta}|n\rangle\right].
\label{Ppm}   
\end{eqnarray}

We note that the $+$ ($-$) part is symmetric (antisymmetric) in both the indices $m\leftrightarrow n$ and $\mu \leftrightarrow \nu$. 
Using the Sokhotski–Plemelj identity, one obtains explicit expressions for the real and imaginary parts of the Kubo correlators in Eq.~\eqref{inter_Lee}:%{\pink [NG: Isolate absorptive part  to simplify things ?]}
\begin{widetext}
\begin{eqnarray}
\notag
    \mathrm{Re}\left[\mathbb{L}_{\alpha\beta}^{\mu\nu}(\omega)\right]&=&\frac{1}{2\hbar A}\sum_{n\neq m}f_{nm}\omega_{mn}\left(2\pi P^{+,\mu\nu}_{\alpha\beta,nm}\delta(\omega-\omega_{mn})-P^{-,\mu\nu}_{\alpha\beta,nm}\,\mathrm{p.v.} \left[\frac{1}{\omega-\omega_{mn}}\right]\right)\,,\\
    \mathrm{Im}\left[\mathbb{L}_{\alpha\beta}^{\mu\nu}(\omega)\right] &=&\frac{1}{2\hbar A}\sum_{n\neq m} f_{nm}\omega_{mn}\left(\pi P^{-,\mu\nu}_{\alpha\beta,nm} \delta(\omega-\omega_{mn})+2 P^{+,\mu\nu}_{\alpha\beta,nm} \,\mathrm{p.v.}\left[\frac{1}{\omega-\omega_{mn}}\right]\right)\,,
    \label{Lmunu_ReIm}
\end{eqnarray}
\end{widetext}
where $\mathrm{p.v.}$ denotes the principal value.

To isolate the bulk contribution and preserve the insulating character free from boundary effects, it is convenient to consider crystalline systems with periodic boundary conditions. In this case, the response can be evaluated in the Bloch basis $|u_{n\bm{k}}\rangle$, with $\bm{k}$ the crystal quasimomentum and where $n$ now represents a band index. In this representation, the off-diagonal matrix elements of the polarization operators are well-defined and given by
\begin{eqnarray}
%\notag
\langle u_{n\bm{k}}|\hat{P}_e^{\mu}|u_{m\bm{k}}\rangle &=& -i e \langle u_{n\bm{k}}|\partial_{k_{\mu}} u_{m\bm{k}}\rangle,\label{pol_nm}\\
\notag
\langle u_{n\bm{k}}|\hat{P}_{Q}^{\mu}|u_{m\bm{k}}\rangle &=& \frac{i}{2}(\varepsilon_{n\bm{k}} + \varepsilon_{m\bm{k}} - 2 \mu)\langle u_{n\bm{k}}|\partial_{k_{\mu}} u_{m\bm{k}}\rangle,
\end{eqnarray}
highlighting their direct connection to the geometric structure of the Bloch wavefunctions.

From Eqs.~\eqref{Lmunu_DC} and~\eqref{Lmunu_ReIm}, one finds that the DC limit of the Kubo correlation functions is governed solely by the spatially antisymmetric component and can be written as
\begin{eqnarray}   
\mathbb{L}_{\alpha\beta}^{\mathrm{DC},\mu\nu}\!\!&=&\!\!\mathrm{Re}\left[\mathbb{L}_{\alpha\beta}^{\mu\nu}(0)\right]\!\!=\!\frac{2}{\pi}\int_{0}^{\infty}\!d\omega\frac{\mathrm{Im}\left[\mathbb{L}_{\alpha\beta}^{\mu\nu}(\omega)\right]}{\omega}\,,
\label{Kramers-Kronig_L}\\
\!\!&=&\!\!\frac{1}{\hbar}\!\!\sum_{n\neq m}\!\int\!\!\!\frac{d^2k}{(2\pi)^2}\!\!\int_{0}^{\infty}\!\!\!\!d\omega f_n P^{-,\mu\nu}_{\alpha\beta,nm}(\bm{k}) \delta(\omega-\omega_{mn}(\bm{k})),\notag
\end{eqnarray}
where
\begin{equation}
P^{-,\mu\nu}_{\alpha\beta,nm} (\bm{k})= 2\mathrm{Im}\left[\langle u_{n\bm{k}}|\hat{P}^{\mu}_{\alpha}|u_{m\bm{k}}\rangle \langle u_{m\bm{k}}|\hat{P}^{\nu}_{\beta}|u_{n\bm{k}}\rangle\right]\label{Pmk}.
\end{equation}
The spectral sum rules in Eq.~\eqref{Kramers-Kronig_L} directly follow from the Kramers–Kronig relations~\cite{bennett1965faraday} and the odd-frequency property $\mathrm{Im}\left[\mathbb{L}_{\alpha\beta}^{\mu\nu}(\omega)\right]=-\mathrm{Im}\left[\mathbb{L}_{\alpha\beta}^{\mu\nu}(-\omega)\right]$. Importantly, these sum rules impose integral constraints on the system's absorption spectrum.

Using Eqs.~\eqref{pol_nm}-\eqref{Pmk}, and taking the zero-temperature limit, the DC Kubo coefficients in the insulating ground state can be compactly written as
\begin{subequations}
\label{L_DC_T0_final}
\begin{align}
\label{L_ee_DC}
\frac{2}{\pi}\int_{0}^{\infty}\!d\omega\frac{\mathrm{Im}\left[\mathbb{L}_{ee}^{\mu\nu}(\omega)\right]}{\omega} =\mathbb{L}_{ee}^{\mathrm{DC},\mu\nu}
&= -\frac{e^2}{h}\,\epsilon^{\mu\nu\rho}C_{\rho},\\
%\label{L_eQ_DC}
\frac{2}{\pi}\int_{0}^{\infty}\!d\omega\frac{\mathrm{Im}\left[\mathbb{L}_{eQ}^{\mu\nu}(\omega)\right]}{\omega}=\mathbb{L}_{eQ}^{\mathrm{DC},\mu\nu}
&= -c\,\epsilon^{\mu\nu\rho}\mathcal{M}_{\rho},\label{magn_sum_rule}\\
\label{L_QQ_DC}
\frac{2}{\pi}\int_{0}^{\infty}\!d\omega\frac{\mathrm{Im}\left[\mathbb{L}_{QQ}^{\mu\nu}(\omega)\right]}{\omega}=\mathbb{L}_{QQ}^{\mathrm{DC},\mu\nu}
&= -2\,\epsilon^{\mu\nu\rho}\mathcal{M}^{Q}_{\rho},
\end{align}
\end{subequations}
where $C_{\rho}$, $\mathcal{M}_{\rho}$ and $\mathcal{M}_{\rho}^{Q}$ denote, respectively, the ground-state Chern number, and the zero-temperature orbital and heat magnetization densities. These quantities are given by the following $\bm{k}$-space integrals~\cite{xiao2005berry,ceresoli2006orbital,shi2007quantum, matsumoto2011rotational, qin2011energy}
\begin{subequations}
\label{defs_CM}
\begin{align}
\label{C_def}
C_{\rho}
&= \frac{i\epsilon_{
\rho\mu\nu
}}{2\pi}\int d^2k \sum_{n\in\mathrm{occ}} \langle\partial_{k_\mu}u_{n\bm{k}}|\partial_{k_\nu} u_{n\bm{k}}\rangle,\\
\label{M_def}
\mathcal{M}_{\rho}
&=\!\frac{\epsilon_{\rho\mu\nu}}{2 i\Phi_0}\!
\int\!\!\frac{d^2k}{2\pi}\!\!\!\sum_{n\in\mathrm{occ}}\langle\partial_{k_\mu}u_{n\bm{k}}
|\hat{H}_{\bm{k}}+\varepsilon_{n\bm{k}}-2\mu|
\partial_{k_\nu}u_{n\bm{k}}\rangle,\\
\label{MQ_def}
\mathcal{M}^{Q}_{\rho}
&=\!\frac{i\epsilon_{\rho\mu\nu}}{2\hbar}
\!\!\!\int\!\!\!\frac{d^2k}{(2\pi)^2}\!\!\!\!\sum_{n\in\mathrm{occ}}
\!\langle\partial_{k_\mu}u_{n\bm{k}}|
\left(\!\!\frac{\hat{H}_{\bm{k}}\!+\!\varepsilon_{n\bm{k}}\!-\!2\mu}{2}\!\!\right)^{2}
\!\!\!|\partial_{k_\nu}u_{n\bm{k}}\rangle,
\end{align}
\end{subequations}
where $\Phi_0\!=\!hc/e$ denotes the flux quantum. 

Equations~\eqref{transp_coeff} and~\eqref{L_DC_T0_final} make explicit that all thermoelectric and thermal transport coefficients vanish in the zero-temperature limit ($\alpha^{\mu\nu}=\overline{\alpha}^{\mu\nu}=\kappa^{\mu\nu}=0$), consistent with the absence of entropy and heat transport required by the third law of thermodynamics. The magnetization subtractions in Eq.~\eqref{transp_coeff} are essential to ensure this behavior and to eliminate unphysical divergences that would otherwise appear in this limit. The only transport response that can remain finite at zero temperature is the electrical Hall conductivity, $\sigma^{\mu\nu}=-e^2 \epsilon^{\mu\nu\rho}C_{\rho}/h$, which remains topologically quantized according to the ground-state Chern number~\cite{niu1985quantized}.

Interestingly, the Kramers–Kronig relations in Eq.~\eqref{L_DC_T0_final} imply that all the ground-state quantities in Eq.~\eqref{defs_CM} admit a spectral representation, establishing a direct link between static ground-state properties and the system's excitation spectrum. While this connection is well-established for the Chern number~\cite{tran2017probing,goldman2024relating}, the fact that the \textit{full} orbital and heat magnetization densities can also be obtained from thermoelectric and thermal correlation functions has been largely overlooked.

For later use, it is convenient to introduce the absorptive (Hermitian) part of the Kubo correlation functions: 
\begin{eqnarray}
\notag
\mathbb{L}^{\mathrm{abs},\mu\nu}_{\alpha\beta}(\omega)\!&=&\!\frac{1}{2}\left(\mathbb{L}^{\mu\nu}_{\alpha\beta}(\omega)+\mathbb{L}^{\nu\mu *}_{\alpha\beta}(\omega)\right)\\
    \notag
    \!&=&\!\frac{\pi}{\hbar}\!\!\int\!\!\frac{d^2k}{(2\pi)^2}\!\!\sum_{n\neq m}f_{nm}(\bm{k})\omega_{mn}(\bm{k})\delta(\omega-\omega_{mn}(\bm{k}))\\
    \!& &\!\times \left(P_{\alpha\beta,nm}^{+,\mu\nu}(\bm{k})+i\frac{P_{\alpha\beta,nm}^{-,\mu\nu}(\bm{k})}{2}\right)\, .
    \label{Labs}
\end{eqnarray}
This quantity represents the component of the response that is genuinely associated with optical absorption. Therefore, it is the quantity that naturally appears in sum rules and experimentally observed transition rates. In particular, we note that the Kramers-Kronig relations in Eqs.~\eqref{Kramers-Kronig_L} and~\eqref{L_DC_T0_final} can equivalently be expressed in terms of $\mathrm{Im}[\mathbb{L}_{\alpha\beta}^{\mathrm{abs},\mu\nu}(\omega)]$.

\subsection{Relation to the dichroic $f$-sum rule}

Dichroic approaches entirely based on the electrical conductivity tensor, such as the dichroic $f$-sum rule~\cite{oppeneer1998magneto,verma2024instantaneous,souza2008dichroic}, only capture \emph{partial} contributions to the orbital magnetization, missing relevant terms that arise naturally from thermoelectric and thermal correlations. Concretely, the dichroic $f$-sum rule~\cite{oppeneer1998magneto,verma2024instantaneous,souza2008dichroic} takes the form
\begin{equation}
    \int_{0}^{\infty}\!d\omega \, \mathrm{Im} \, \left[\mathbb{L}_{ee}^{\mathrm{abs},\mu\nu}(\omega)\right]= \epsilon_{\mu\nu\rho} \frac{\pi e c}{\hbar} \mathcal{M}_\rho^{{\rm SR}},\label{f_sum_rule}
\end{equation}
%\begin{equation}
 %   \int_{0}^{\infty}\!d\omega \, \mathrm{Im} \, \left[\sigma^{\mu\nu}(\omega)\right]= (\dots) \epsilon_{\mu\nu\rho} \mathcal{M}_\rho^{{\rm SR}},\label{f_sum_rule}
%\end{equation}
where the ground-state property
\begin{equation}
\mathcal{M}^{{\rm SR}}_\rho= \frac{1}{2i\Phi_0}\epsilon_{\rho \mu\nu}\!\!\int\!\!\frac{d^2k}{2\pi}\!\! \sum_{n\in\mathrm{occ}}\braket{\frac{\partial u_{n\bm{k}}}{\partial k_\mu}|\hat{H}_{\bm{k}}-\varepsilon_{n\bm{k}}|\frac{\partial u_{n\bm{k}}}{\partial k_\nu}}\, ,
\end{equation}
accounts for the intrinsic magnetic-moment (or \emph{self-rotation}) contribution of the Bloch states to the orbital magnetization~\cite{chang1996berry,xiao2005berry,xiao2010berry}. 

It is well known that the total orbital magnetization of Bloch bands can be expressed in terms of two contributions~\cite{chang1996berry,xiao2010berry,matsumoto2011rotational}, 
\begin{equation}
\bm{\mathcal{M}}\!=\!\bm{\mathcal{M}}^{{\rm SR}}+\bm{\mathcal{M}}^{{\rm COM}}, \label{two_contributions}
\end{equation}
where the \emph{center-of-mass} contribution
\begin{equation}
    \mathcal{M}_{\rho}^{{\rm COM}}=-\frac{1}{\Phi_0}\int \frac{d^2k}{2\pi} \sum_{n\in\mathrm{occ}} \left(\varepsilon_{n\bm{k}}-\mu\right)\Omega_{n\bm{k}}^\rho\,,\label{orb_COM}
\end{equation}
involves the Berry curvature of the occupied bands
\begin{equation}    \Omega_{n\bm{k}}^{\rho}=i\epsilon_{\rho\mu\nu}\braket{\frac{\partial u_{n\bm{k}}}{\partial k_\mu}|\frac{\partial u_{n\bm{k}}}{\partial k_\nu}}. 
\end{equation}

We thus conclude that this center-of-mass contribution to the orbital magnetization, which is absent in the dichroic $f$-sum rule in Eq.~\eqref{f_sum_rule}, is captured by thermoelectric correlations in the generalized sum rule~\eqref{magn_sum_rule}. 

As a corollary, we conclude that the two distinct contributions to the orbital magnetization in Eq.~\eqref{two_contributions} can be individually measured, by combining the two sum rules in Eqs.~\eqref{f_sum_rule} and \eqref{magn_sum_rule}. A similar conclusion can be reached for the heat magnetization, as we further discuss in Section~\ref{section_generalization}.\\

In the following Section~\ref{Section_generalized_dichroism}, we will show how the absorptive part of the frequency-resolved Kubo response functions can be accessed experimentally by monitoring excitation rates under suitably engineered thermoelectric probes, providing a direct and practical route to extract the full set of ground-state properties discussed above.

%{\pink [NG: I wonder how to improve the introduction/definition of \emph{the absorptive part} in the manuscript.]}

\subsection{The heat quantum metric}

Interestingly, the sum rules in Eqs.~\eqref{Kramers-Kronig_L} and \eqref{L_DC_T0_final} also admit a real-part analogue. In the zero-temperature limit, this reads
\begin{eqnarray}
\notag
        \int_{0}^{\infty}\!\!\!d\omega \frac{\mathrm{Re}\left[\mathbb{L}_{\alpha\beta}^{\mathrm{abs},\mu\nu}(\omega)\right]}{\omega}\!&=&\!\frac{\pi}{\hbar}\!\int\!\!\frac{d^2k}{(2\pi)^2}\!\!\!\sum_{n \in \textrm{occ}}\sum_{m\in \textrm{unocc}}\!\!\!\!P^{+,\mu\nu}_{\alpha\beta,nm}(\bm{k})\\
        &\equiv& \frac{\pi}{\hbar}\int\!\!\frac{d^2k}{(2\pi)^2}g_{\alpha\beta}^{\mu\nu}(\bm{k}).
    \label{SWM_generalized}
\end{eqnarray}
This expression isolates the real components of the matrix elements in Eq.~\eqref{P_decomp}, thereby defining the symmetric tensors 
\begin{eqnarray}
\notag
    g_{\alpha\beta}^{\mu\nu}(\bm{k})\!&=&\!\! \sum_{n\in\textrm{occ}}\sum_{m\in\textrm{unocc}}\!\!\mathrm{Re}\left[\langle u_{n\bm{k}}|\hat{P}_{\alpha}^{\mu}|u_{m\bm{k}}\rangle\langle u_{m\bm{k}}|\hat{P}^{\nu}_{\beta}|u_{n\bm{k}}\rangle\right]\\
    &=&\hbar^2\sum_{\substack{n\in\textrm{occ}\\m\in\textrm{occ}}}\mathrm{Re} \,
\frac{
\langle u_{n\bm{k}} | \hat{J}_{\alpha}^{\mu} | u_{m\bm{k}} \rangle
\langle u_{m\bm{k}} | \hat{J}_{\beta}^{\nu} | u_{n\bm{k}} \rangle
}{
(\varepsilon_{n\bm{k}} - \varepsilon_{m\bm{k}})^2 
}.
\end{eqnarray}
One readily recognizes the well-known Souza-Wilkens-Martin sum rule~\cite{souza2000polarization} by taking $\alpha\!=\!\beta\!=\!e$ in Eq.~\eqref{SWM_generalized}, and noting that $g_{ee}^{\mu\nu}(\bm{k})\!=\!e^2 g^{\mu\nu}(\bm{k})$, with $g^{\mu\nu}(\bm{k})$ the quantum metric associated with the occupied Bloch bands~\cite{kolodrubetz2017geometry,yu2024quantum,ozawa2019probing}.

%Furthermore, we note that Eq.~\eqref{SWM_generalized} is directly related to the quadratic quantum fluctuations of the corresponding polarization operators in the ground state,
%\begin{equation}
 %   \int \frac{d^2k}{(2\pi)^2}g_{\alpha\beta}^{\mu\nu}(\bm{k})\!=\!\frac{1}{A}\!\left(\!\frac{\langle \hat{P}_{\alpha}^{\mu}\hat{P}_{\beta}^{\nu} +\hat{P}_{\beta}^{\nu}\hat{P}_{\alpha}^{\mu}\rangle_0}{2}-\langle \hat{P}_{\alpha}^{\mu}\rangle_0\langle\hat{P}_{\beta}^{\nu}\rangle_0\!\right),\label{eq_fluctuations}
%\end{equation}
%where $\langle \dots \rangle_0$ denotes the ground-state expectation value. 

Considering the thermal case, $\alpha\!=\!\beta\!=\!Q$, we obtain the symmetric tensor
\begin{eqnarray}
\notag
    g_{QQ}^{\mu\nu}(\bm{k})\!&=&\!\sum_{n\in \textrm{occ}}\, \sum_{m\textrm{unocc}} \mathrm{Re}\left[\langle u_{n\bm{k}}|\hat{P}_{Q}^{\mu}|u_{m\bm{k}}\rangle\langle u_{m\bm{k}}|\hat{P}_Q^{\nu}|u_{n\bm{k}}\rangle\right]\\
    \notag
    \!&=&\!\sum_{n\in \textrm{occ}}\,\sum_{m\textrm{unocc}}\left(\frac{\varepsilon_{n\bm{k}}+\varepsilon_{m\bm{k}}-2\mu}{2}\right)^2\\
    & &\times\mathrm{Re}\left[\langle \partial_{k_{\mu}}u_{n\bm{k}}|u_{m\bm{k}}\rangle\langle u_{m\bm{k}}|\partial_{k_{\nu}}u_{n\bm{k}}\rangle\right].\label{eq_heat_metric}
\end{eqnarray}
We identify $g_{QQ}^{\mu\nu}$ as a ``heat quantum metric", noting that this symmetric tensor is semi-positive defined with respect to the scalar product in the Cartesian indices. The geometric meaning of this heat quantum metric can be understood in close analogy with the standard quantum metric, as we now explain.

In general, for a family of Bloch Hamiltonians $\hat H_{\bm k} (\bm{\lambda})$ depending on a set of parameters $\bm{\lambda}$, one can define a $\bm{k}$-resolved quantum metric~\cite{kolodrubetz2017geometry}
\begin{equation}
g^{\mu\nu}_{\bm{\lambda}}(\bm{k}) 
\!\!=\!\!\!\!\sum_{\substack{n \in \mathrm{occ}\\m \in \mathrm{unocc}}}\!\!\!\!
\mathrm{Re} \,
\frac{
\langle u_{n\bm{k}} | \partial_{\lambda_\mu} \hat{H}_{\bm{k}} | u_{m\bm{k}} \rangle
\langle u_{m\bm{k}} | \partial_{\lambda_\nu} \hat{H}_{\bm{k}} | u_{n\bm{k}} \rangle
}{
(\varepsilon_{n\bm{k}} - \varepsilon_{m\bm{k}})^2
},
\end{equation}
which measures the infinitesimal distance between the occupied subspaces defined by the projector $\mathcal{P}_k = \sum_{n\in \textrm{occ}} |u_{n\bm{k}}\rangle\langle u_{n\bm{k}}|$ at nearby values of $\bm{\lambda}$. Equivalently, it can be written as $g^{\mu\nu}_{\bm{\lambda}}(\bm{k}) = \frac{1}{2}\mathrm{Tr}[(\partial_{\lambda_\mu}\mathcal{P}_{\bm{k}})(\partial_{\lambda_\nu}\mathcal{P}_{\bm{k}})]$, making explicit its interpretation as a distance on the manifold of projectors onto occupied bands. The physical content of the metric $g^{\mu\nu}_{\bm{\lambda}}(\bm{k})$ depends on the choice of parameters $\bm{\lambda}$. For instance, choosing
$\lambda_e^{\mu} = \frac{e}{\hbar c} A_e^\mu$
with $\mathbf{A}_e$ a constant electromagnetic vector potential that couples to the charge current, one recovers the usual quantum metric $g^{\mu\nu}_{\bm{\lambda}_e}(\bm{k}) = g^{\mu\nu}(\bm{k})$, since $\partial_{\lambda_e^\mu}\hat{H}_{\mathbf{k}} = \partial_{k^\mu}\hat{H}_{\mathbf{k}}$. This construction is equivalent to introducing twists of the boundary conditions, i.e., threading electromagnetic fluxes through the system. In analogy, the heat quantum metric can be obtained by considering a gravitomagnetic vector potential $\mathbf{A}_g$ coupling to the heat current~\cite{Tatara2015,zhang2020thermodynamics,Buthenhoff2026}, such that $\partial_{{A_g}_\mu}\hat{H}_{\mathbf{k}} = -\hat{J}_Q^\mu(\mathbf{k})$. Choosing $\lambda_Q^\mu = \frac{1}{\hbar} A_g^\mu$, the corresponding metric defines the heat quantum metric $g^{\mu\nu}_{\bm{\lambda}_Q}(\bm{k})=g_{QQ}^{\mu\nu}(\bm{k})$ in Eq.~\eqref{eq_heat_metric}, which can be viewed as an energy-weighted analogue of the usual Bloch quantum metric. In direct analogy with the electromagnetic case, the deformations generated by $\lambda_Q^\mu$ correspond to twists of the gravitomagnetic potential, playing a role analogous to boundary-condition phases associated with electromagnetic fluxes. Interestingly, within finite-temperature formulations, such deformations can be understood more explicitly as twists in a gravitational (thermal) flux space~\cite{Nakai2017,Nakai2022}. From this perspective, the integrated metric $\int d^2 k g_{QQ}^{\mu\nu}(\bm{k})$, which dictates the response of a filled Bloch band in Eq.~\eqref{SWM_generalized}, can be interpreted as a \emph{many-body heat quantum metric}~\cite{ozawa2019probing,Faugno_MBQM} defined over the parameter space associated with these thermal twists.

We note that the heat quantum metric in Eq.~\eqref{eq_heat_metric} has recently been shown to enter physical observables: in particular, it contributes to the thermal Meissner stiffness~\cite{Buthenhoff2026}, much in the same way that the usual quantum metric contributes to the superfluid stiffness. In Section~\ref{section_open_exp}, we use a real‑space representation to interpret the relation in Eq.~\eqref{SWM_generalized} in the context of the fluctuation–dissipation theorem for the heat‑polarization operator $\hat P_{\alpha}^{\mu}$.

%{\pink We note that more exotic hybrid quantum metrics $g_{eQ}^{\mu\nu}$ could  be considered~\cite{shinada2025quantum}.}

\section{Thermoelectric dichroism and spectral sum rules from integrated differential rates}\label{Section_generalized_dichroism}

In this Section, we show that the dichroic approach introduced in Refs.~\cite{tran2017probing,repellin2019detecting,asteria2019measuring,goldman2024relating} and reminded in Section~\ref{section_illustration} -- originally proposed to extract the many-body Chern number of an insulator by monitoring Bloch-band excitation rates under a circular drive -- can be naturally extended to access the full equilibrium orbital and heat magnetization densities.

\subsection{Thermoelectric probes and sum rules}

We consider a crystalline lattice subjected to a chiral thermoelectric drive of the form 
\begin{equation}
    \delta\hat{H}_{\alpha\beta}^{\pm,\mu\nu}(t)=\frac{2}{\omega} \left(V_{0\alpha}\sin(\omega t)\hat{J}_{\alpha}^{\mu} \mp V_{0\beta}\cos(\omega t)\hat{J}^{\nu}_{\beta}\right),\label{chiral_th_probe_eq}
\end{equation}
where the dimensional prefactors $V_{0\alpha}$ are chosen so that lowest-order time-dependent perturbation theory remains applicable. As in Section~\ref{section_sum_rules}, the indices $\alpha,\beta\!=\!e,Q$ specify the nature of the thermoelectric components entering the probe, hence providing a direct generalization of the electric probe in Eq.~\eqref{eq:perturbation}. Here $V_{0e}$ carries units of electric field, while $V_{0Q}$ carries units of inverse length.

Following Fermi's golden rule, and considering a fermionic system at zero temperature, the excitation rate from occupied to unoccupied Bloch states for a driving frequency $\omega$ can be obtained as
\begin{eqnarray}
\notag   \Gamma_{\alpha\beta}^{\pm,\mu\nu}(\omega)&=&\frac{2\pi A}{\hbar^2}\!\!\int \!\!\frac{d^2k}{(2\pi)^2}\!\!\sum_{n\in \textrm{occ}}\,\sum_{m\in \textrm{unocc}}\!\!\frac{\delta(\omega-\omega_{mn}(\bm{k}))}{\omega^2}\\
    & &\times  |\langle u_{m\bm{k}}|V_{0\alpha}\hat{J}_{\alpha}^{\mu}\pm i V_{0\beta}\hat{J}_{\beta}^{\nu}|u_{n\bm{k}}\rangle|^2.
    \label{Gamma_pm_PBC}
\end{eqnarray}

\begin{figure*}[t]
    \centering
    \includegraphics[width=0.75\textwidth]{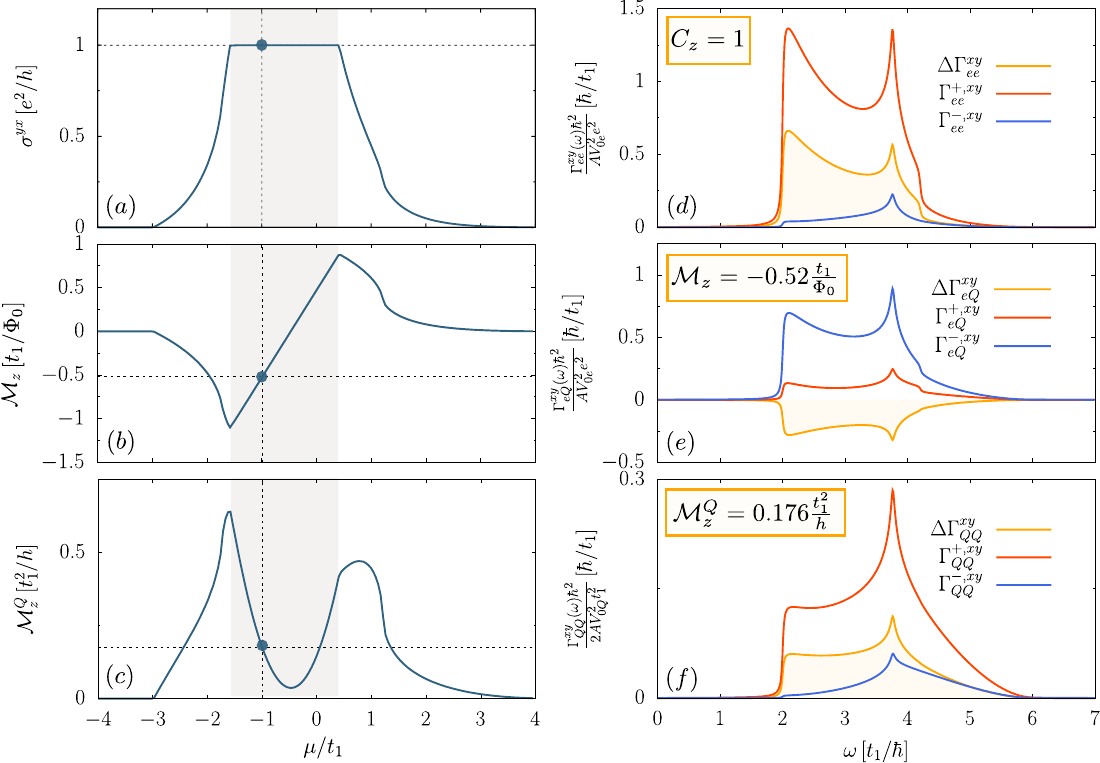}
    \caption{Panels $(a)$, $(b)$, and $(c)$ show the Hall conductivity, orbital magnetization, and heat magnetization of the Haldane model as a function of the chemical potential $\mu$, for $\Delta_{AB}=0$, $t_2/t_1=0.5$, and $\varphi=0.7\pi$. The gray shaded region indicates the bulk gap. Panels $(d)$, $(e)$, and $(f)$ display the frequency-resolved excitation rates $\Gamma_{\alpha\beta}^{\pm,xy}(\omega)$ in Eq.~\eqref{Gamma_pm_PBC} and the corresponding differential rates $\Delta\Gamma_{\alpha\beta}^{xy}(\omega)$ (see Eq.~\eqref{Delta_Gamma_PBC}) for the three chiral drives $ee$, $eQ$, and $QQ$. For the mixed electric-heat drive ($eQ$) we have chosen $V_{0e}/V_{0Q} = t_1/e$. The delta functions in the rate expressions have been approximated by Lorentzians with a broadening of $\eta/t_1=0.02$ to improve visibility. The chemical potential is $\mu/t_1=-1$ (filled circle in panels $(a)$-$(c)$). The integrated differential rate is indicated in the boxed inset.}
    \label{rates_PBC}
\end{figure*}

The difference between excitation rates obtained for right ($+$) and left ($-$) handed circular perturbations defines the differential rate
\begin{equation}
 \Delta \Gamma_{\alpha\beta}^{\mu\nu}(\omega)=\frac{1}{2}\left(\Gamma_{\alpha\beta}^{+,\mu\nu}(\omega)-\Gamma_{\alpha\beta}^{-,\mu\nu}(\omega)\right),   
 \label{Delta_Gamma_PBC}
\end{equation}
which evaluates to

\begin{eqnarray}
\notag
    \frac{\Delta \Gamma^{\mu\nu}_{\alpha\beta}(\omega)}{A}\!&=&\!-\frac{2\pi}{\hbar^2}\!\!\int\!\!\frac{d^2k}{(2\pi)^2}\!\!\sum_{n\in\textrm{occ}}\,\sum_{m\in\textrm{unocc}}\!\!\!V_{0\alpha}V_{0\beta}\\
    & &\times P_{\alpha\beta,nm}^{-,\mu\nu}(\bm{k})\delta(\omega-\omega_{mn}(\bm{k})).
    \label{DR}
\end{eqnarray}
We note that, while the frequency-resolved excitation rates $\Gamma_{\alpha\beta}^{\pm,\mu\nu}(\omega)$ generally depend on the lattice embedding (i.e.~the relative position of orbitals), the differential rates $\Delta\Gamma^{\mu\nu}_{\alpha\beta}(\omega)$ are independent of the lattice geometry~\cite{Simon2020}.

Combining Eq.~\eqref{DR} with the definition of the absorptive correlator in Eq.~\eqref{Labs} yields, for $\omega>0$,
\begin{equation}
    \mathrm{Im}\left[\mathbb{L}_{\alpha\beta}^{\mathrm{abs},\mu\nu}(\omega)\right]=-\frac{\hbar\omega\Delta\Gamma_{\alpha\beta}^{\mu\nu}(\omega)}{4 A V_{0\alpha}V_{0\beta}}.
    \label{ImLabs_rates}
\end{equation}
Equation~\eqref{ImLabs_rates}, which generalizes Eq.~\eqref{eq:power_absorption} to thermoelectric responses, is a remarkable relation: it shows that the imaginary part of the absorptive Kubo tensor can be directly extracted from the frequency-resolved differential rates measured under circularly polarized (chiral) drives.

Combining Eqs.~\eqref{Kramers-Kronig_L} and~\eqref{DR}, one finds a direct correspondence between the integrated differential rates  and the DC Kubo coefficients in Eq.~\eqref{L_DC_T0_final}:
\begin{eqnarray}
\notag
    \frac{\Delta\Gamma^{\textrm{int},\mu\nu}_{\alpha\beta}}{A} &=& \int_{0}^{\infty}d\omega \frac{\Delta \Gamma^{\mu\nu}_{\alpha\beta}(\omega)}{A}\\
    &=& -\frac{2\pi}{\hbar}V_{0\alpha}V_{0\beta} \mathbb{L}_{\alpha\beta}^{\mathrm{DC},\mu\nu},
    \label{DIR}
\end{eqnarray}
such that
\begin{subequations}
    \begin{align}
       \Delta\Gamma^{\textrm{int},\mu\nu}_{ee}/A &= V_{0e}^2 \frac{e^2}{\hbar^2}\epsilon_{\mu\nu\rho}C_{\rho}\,, \\
       \Delta\Gamma^{\textrm{int},\mu\nu}_{eQ}/A &= V_{0e}V_{0Q}\frac{2\pi c}{\hbar}\epsilon_{\mu\nu\rho}\mathcal{M}_{\rho} \,,\label{eq_dichroism_orbital}\\
       \Delta\Gamma^{\textrm{int},\mu\nu}_{QQ}/A &= \frac{4\pi}{\hbar}V_{0Q}^2\epsilon_{\mu\nu\rho}\mathcal{M}_{\rho}^{Q} \label{eq_dichroism_heat}.
    \end{align}
    \label{DIR2}
\end{subequations}

The frequency-integrated differential rates therefore act as direct experimental probes for the corresponding DC Kubo coefficients of insulators, linking measurable excitation asymmetries (i.e.~generalized circular dichroism) to intrinsic ground-state properties. In particular, purely electrical drives yield the Chern number, while mixed and purely thermal drives ($eQ$ and $QQ$) respectively give access to the full orbital and heat magnetization densities. This establishes a unified framework in which topological and magnetization properties emerge on equal footing from integrated dichroic spectra obtained under appropriately designed chiral drives.\\

\subsection{Validation on the Haldane model}

To illustrate these relations and their physical content, we now turn to a concrete example. We consider the Haldane model~\cite{haldane1988model,asteria2019measuring}, a paradigmatic Chern insulator defined on a honeycomb lattice with a sublattice potential offset $\Delta_{AB}$, real nearest-neighbor hopping $t_1$, and complex next-nearest-neighbor hopping $t_2 e^{\pm i \varphi}$ that breaks time-reversal-symmetry without net magnetic flux. Using this model, we compute the frequency-resolved excitation rates $\Gamma^{\pm,xy}_{\alpha\beta}(\omega)$ and the corresponding differential rates $\Delta\Gamma^{xy}_{\alpha\beta}(\omega)$ for the three chiral drives ($ee$, $eQ$, and $QQ$). 

Figure~\ref{rates_PBC} summarizes the results. Panels $(a)$-$(c)$ show the Hall conductivity $\sigma^{yx}$, the orbital magnetization $\mathcal{M}_z$, and the heat magnetization $\mathcal{M}_z^{Q}$ as a function of the chemical potential $\mu$. Within the gapped (gray-shaded) region, the Hall conductivity remains topologically quantized according to the Chern number $C_z\!=\!1$, the orbital magnetization grows linearly with $\mu$ in accordance with  the St\v{r}eda relation~\cite{Auerbach2024}, $\partial \mathcal{M}_z/\partial \mu \!=\!\partial n/\partial B_z\!=\!C_z e/hc$, and the heat magnetization varies quadratically with $\mu$. The quadratic dependence of $\mathcal{M}^Q_z$ inside the gap is also fixed by the Chern number of the occupied bands, since $\partial^2 \mathcal{M}^Q_z/\partial \mu^2= (2c/e)\partial \mathcal{M}_z/\partial \mu=2C_z/h$, as obtained by Eqs.~\eqref{M_def} and~\eqref{MQ_def}.
Panels $(d)$-$(f)$ show the corresponding frequency-resolved rates for the three chiral drives, as calculated for a system with periodic boundary conditions. The integrated values of the differential rates (boxed insets) match the corresponding ground-state properties extracted in panels $(a)$-$(c)$, thereby illustrating the applicability of our approach.

\subsection{Linear drives and the heat quantum metric}\label{sect_linear_drive}

Before concluding this section, we note that the real part of $\mathbb{L}_{\alpha\beta}^{\mathrm{abs},\mu\nu}(\omega)$ can be accessed by using linearly polarized drives~\cite{ozawa2018extracting,ozawa2019probing,yu2020experimental},
\begin{equation}
    \delta\hat{H}_{\alpha\beta}^{\pm,\mu\nu}(t)=\frac{2}{\omega} \left(V_{0\alpha}\hat{J}_{\alpha}^{\mu} \mp V_{0\beta}\hat{J}^{\nu}_{\beta}\right)\sin(\omega t).\label{linear_th_probe_eq}
\end{equation}
Such drives would isolate the symmetric sector of the matrix elements $P_{\alpha\beta,nm}^{+,\mu\nu}$, allowing one to reconstruct $\mathrm{Re}[\mathbb{L}^{\mathrm{abs},\mu\nu}_{\alpha\beta}(\omega)]$, %and the ground-state polarization fluctuations in {\pink %Eqs.~\eqref{SWM_generalized} and~\eqref{eq_fluctuations},} 
from the corresponding (non-chiral) transition rates. 

For instance, measuring the excitation rates $\Gamma_{Q}^{x}(\omega) $ resulting from the linear drive $\delta \hat{H}_{Q}^{x}=2V_{0Q}\hat{J}^{x}_{Q}\cos(\omega t)/\omega$ would yield the average of the diagonal component of the heat quantum metric introduced in Eq.~\eqref{eq_heat_metric}, namely,
\begin{equation}
    \frac{\hbar^2}{2\pi A V_{0Q}^2}\int_{0}^{\infty}d\omega\,\Gamma_{QQ}^{x}(\omega) = \int \frac{d^2k}{(2\pi)^2}g_{QQ}^{xx}(\bm{k}).\label{fluctuation_dissipation}
\end{equation}

The other components of this integrated heat quantum metric could be obtained by varying the orientation of the linear drive~\cite{ozawa2018extracting,ozawa2019probing,yu2020experimental}; see also Section~\ref{section_open_exp}. 

Finally, we point out that the momentum-resolved heat quantum metric $g_{QQ}^{\mu \nu}(\bm{k})$ could be probed locally in $k$-space, by applying this scheme to Bloch wave packets~\cite{ozawa2018extracting}.

%{\pink We point out that Eq.~\eqref{fluctuation_dissipation} reflects the fluctation-dissipation theorem~\cite{kubo1957statistical}, relating the variance of the heat polarization (i.e.~the many-body heat quantum metric) to its dynamical susceptibility (i.e.~the heating rate associated with a thermal probe); see the next Section.}

%{\pink Figure X(a) displays the behavior of the three key quantities $C$, $\mathcal{M}$ and $\mathcal{M}^Q$ as a function of a system parameters. Figure X(b) illustrates the frequency-resolved different rates, as calculated for a system with periodic boundary conditions. The integrated values (inset) correspond to the target ground-state properties, hence validating our approach.}\\

%\textcolor{red}{Include excitation rate plots for Haldane model here}

%{\pink [NG: Insert large figure showing: three quantities in the Haldane model as a function of a parameter; the PBC differential rates as a function of frequency; same for a finite sample with OBC.]}

\section{Open boundary systems:~Local markers and experimental implementations}\label{section_open_exp}

The formal relations between differential excitation rates and the DC Kubo coefficients derived in Section~\ref{Section_generalized_dichroism} rely on periodic boundary conditions, which ensure that the excitation processes involve only bulk transitions between extended Bloch states. In experimentally relevant settings, however, systems necessarily have open boundaries, where localized edge states may appear within the bulk gap and contribute to the excitation rates~\cite{tran2017probing,unal2024quantized}. Understanding how these boundary contributions modify the dichroic excitation spectra is essential to connect this theoretical formulation with measurable observables.\\

\subsection{Real-space formulation and local markers}

A natural way to address this problem is to reformulate the chiral thermoelectric driving protocol [Eq.~\eqref{chiral_th_probe_eq}] directly in terms of the electric and heat polarization operators
\begin{equation}
     \delta\hat{H}_{\alpha\beta}^{\pm,\mu\nu}(t)=2 \left(V_{0\alpha}\cos(\omega t)\hat{P}_{\alpha}^{\mu} \pm V_{0\beta}\sin(\omega t)\hat{P}^{\nu}_{\beta}\right),\label{general_probe}
\end{equation}
hence generalizing the chiral probe in Eq.~\eqref{eq:perturbation_one} to the present thermoelectric framework. According to Fermi's golden rule, the frequency-resolved rates then read
\begin{eqnarray}
\notag
    \Gamma_{\alpha\beta}^{\pm,\mu\nu}(\omega) &=& \frac{2\pi}{\hbar^2}\sum_{n\in\textrm{occ}}\sum_{m\in\textrm{unocc}}\delta(\omega-\omega_{mn})\\
    & &|\langle m|V_{0\alpha}\hat{P}^{\mu}_{\alpha}\pm iV_{0\beta}\hat{P}_{\beta}^{\nu}|n\rangle|^2,
     \label{Gamma_pm_OBC}
\end{eqnarray}
with the corresponding differential rates yielding
\begin{eqnarray}
\notag
    \frac{\Delta \Gamma^{\mu\nu}_{\alpha\beta}(\omega)}{ V_{0\alpha}V_{0\beta}}\!&=&\!-\frac{2\pi}{\hbar^2}\sum_{n\in\textrm{occ}}\sum_{m\in\textrm{unocc}}P_{\alpha\beta,nm}^{-,\mu\nu}\delta(\omega-\omega_{mn}).\\
    \label{DR_OBC}
\end{eqnarray}
The antisymmetric components $P_{\alpha\beta,nm}^{-,\mu\nu}$ are here entirely expressed in their real-space representation, see Eqs.~\eqref{pol_real} and~\eqref{Ppm}.  The corresponding integrated differential rates take the form
\begin{subequations}
    \begin{align}
       \frac{\Delta\Gamma^{\textrm{int},\mu\nu}_{ee}}{V_{0e}^2} &= \frac{e^2}{\hbar^2}\epsilon_{\mu\nu\rho}\int d^2r\,\mathfrak{C}_{\rho}(\bm{r}) \\
       \frac{\Delta\Gamma^{\textrm{int},\mu\nu}_{eQ}}{V_{0e}V_{0Q}} &= \frac{2\pi c}{\hbar}\epsilon_{\mu\nu\rho}\int d^2r\,\mathfrak{M}_{\rho}(\bm{r})\label{magn_real}\\
       \frac{\Delta\Gamma^{\textrm{int},\mu\nu}_{QQ}}{V_{0Q}^2} &= \frac{4\pi}{\hbar}\epsilon_{\mu\nu\rho}\int d^2r\,\mathfrak{M}^Q_{\rho}(\bm{r}),
    \end{align}
\end{subequations}
where $ \mathfrak{C}_{\rho}(\bm{r})$,  $\mathfrak{M}_{\rho}(\bm{r})$ and $\mathfrak{M}_{\rho}^{Q}(\bm{r})$ denote, respectively, the local markers for the Chern, orbital and heat magnetizations:
\begin{widetext}
    \begin{subequations}
   \begin{align}
       \mathfrak{C}_{\rho}(\bm{r}) =& 2\pi i \epsilon_{\rho\mu\nu}\langle \bm{r}|\hat{\mathcal{P}}\hat{r}^{\mu}\hat{\mathcal{Q}}\hat{r}^{\nu}\hat{\mathcal{P}}|\bm{r}\rangle\,,\\
       \mathfrak{M}_{\rho}(\bm{r}) =& \frac{\pi \epsilon^{\rho\mu\nu}}{i\Phi_0} \langle\bm{r}|\hat{\mathcal{P}}\hat{r}^{\mu}\hat{\mathcal{Q}}\hat{H}\hat{\mathcal{Q}}\hat{r}^{\nu}\hat{\mathcal{P}}-\hat{\mathcal{Q}}\hat{r}^{\mu}\hat{\mathcal{P}}\hat{H}\hat{\mathcal{P}}\hat{r}^{\nu}\hat{\mathcal{Q}}|\bm{r}\rangle+\frac{\mu}{\Phi_0}\mathfrak{C}_{\rho}(\bm{r})\,,\\
       \mathfrak{M}_{\rho}^{Q}(\bm{r}) =& \frac{i \epsilon^{\rho\mu\nu}}{8\hbar}\langle \bm{r}|2\hat{H}\hat{\mathcal{P}}\hat{r}^{\mu}\hat{\mathcal{Q}}\hat{H}\hat{\mathcal{Q}}\hat{r}^{\nu}\hat{\mathcal{P}} + \hat{\mathcal{P}}\hat{r}^{\mu}\hat{\mathcal{Q}}\hat{H}^2\hat{\mathcal{Q}}\hat{r}^{\nu}\hat{\mathcal{P}}-\hat{\mathcal{Q}}\hat{r}^{\mu}\hat{\mathcal{P}}\hat{H}^2\hat{\mathcal{P}}\hat{r}^{\nu}\hat{\mathcal{Q}}|\bm{r}\rangle + \mu\frac{c}{e}\mathfrak{M}_{\rho}(\bm{r}) - \frac{\mu^2}{4\pi \hbar}\mathfrak{C}_{\rho}(\bm{r})\,. 
   \end{align} 
   \label{local_markers}
\end{subequations}
\end{widetext}
Here $\hat{\mathcal{P}} = \sum_{\varepsilon_n<\mu}|n\rangle\langle n|$ is the ground-state projector and $\hat{\mathcal{Q}}=\hat{\mathcal{I}}-\hat{\mathcal{P}}$ its complement. Notably, in this real-space formulation, the familiar markers for the Chern and orbital magnetizations~\cite{Bianco2011,Bianco2013,Resta_request,seleznev2023towards} emerge naturally, together with an analogous expression for the heat magnetization, which, to our knowledge, has not been previously reported. Thanks to the exponential locality of $\hat{\mathcal{P}}(\bm{r},\bm{r}')$ in insulators, the bulk-averaged values of these markers coincide with those on the torus in the thermodynamic limit.  Importantly, while the total Chern number vanishes in an open boundary sample~\cite{Bianco2011,Bianco2013,tran2017probing,tran2018quantized} -- as a direct consequence of the bulk-edge correspondence~\cite{unal2024quantized} -- the total orbital and heat magnetizations remain finite and correspond to their value on the torus~\cite{Wang2022}.

%{\pink[NG: Local edge spectroscopy to isolate edge contribution ? Spectral and spatial resolution for magnetization ? Simulations ? Result from pure edge theory (\`a la Wen) ?]}
\begin{figure}[t]
    \centering
    \includegraphics[width=0.92\columnwidth]{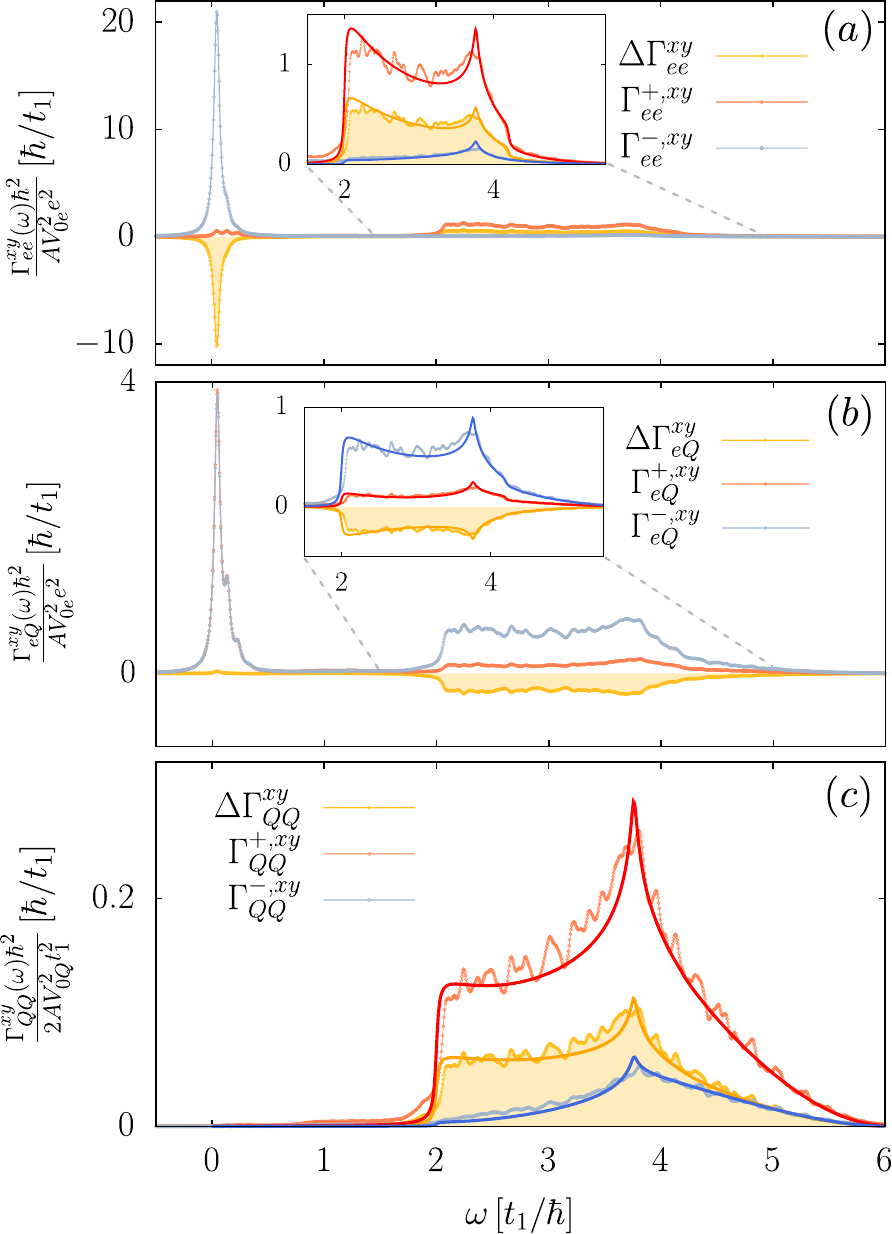}
    \caption{Panels $(a)$, $(b)$ and $(c)$ show the frequency-resolved excitation rates $\Gamma_{\alpha\beta}^{\pm,xy}(\omega)$ [Eq.~\eqref{Gamma_pm_OBC}] and the corresponding differential rates $\Delta\Gamma_{\alpha\beta}^{xy}(\omega)$   [Eq.~\eqref{DR_OBC}] for the three chiral drives ($ee$, $eQ$, and $QQ$) in a Haldane-model flake with open boundaries. The system contains $15 \times 60$ sites, and all model parameters match those used in Fig.~\ref{rates_PBC}$(a)$-$(c)$. The insets in panels $(a)$-$(b)$ zoom in on the bulk response. Thick solid lines indicate the corresponding thermodynamic-limit rates obtained with periodic boundary conditions. For the mixed electric-heat drive ($eQ$), we set $V_{0e}/V_{0Q} = t_1/e$. Delta functions in the rate expressions are approximated by Lorentzians with broadening $\eta/t_1=0.03$ to improve visibility.}
    \label{rates_OBC}
\end{figure}

Figure~\ref{rates_OBC} displays the frequency-resolved excitation rates in a finite open-boundary rectangular flake of $15\times 60$ sites realizing the Haldane model, for the three different types of chiral drives. Thick solid lines indicate the corresponding thermodynamic-limit rates obtained with periodic boundary conditions, extracted from Fig.~\ref{rates_PBC}$(a)$-$(c)$, which nicely compare with the finite-flake response at frequencies above the bulk gap. As anticipated, the differential rate for the purely electrical drive, $\Delta\Gamma_{ee}^{xy}(\omega)$ in Fig.~\ref{rates_OBC}$(a)$,  integrates to zero: the low-frequency edge contribution precisely cancels the integrated bulk response~\cite{unal2024quantized}. In contrast, the frequency-resolved differential rates for the electric-heat and heat-heat drives, $\Delta\Gamma_{eQ}^{xy}(\omega)$ and $\Delta\Gamma_{QQ}^{xy}(\omega)$, are dominated entirely by bulk contributions. This immediately implies that the local markers for the orbital and heat magnetizations in Eq.~\eqref{local_markers} integrate to zero over the edge of the sample, or equivalently, that their spatial average over the entire flake coincides with their spatial average over the bulk region, consistent with previous observations~\cite{Bianco2013,Wang2022}. From the integrated differential rates in Figs.~\ref{rates_OBC}$(b)$ and $(c)$, we extract orbital and heat-magnetizations of approximately $\mathcal{M}_z \simeq -0.53\,t_1/\Phi_0$ and $\mathcal{M}_z^{Q}\simeq 0.18 t_1^2/h$, respectively, which slightly differ from their thermodynamic values [Fig.~\ref{rates_PBC}] due to finite size effects.

\subsection{Linear drives: Fluctuation-dissipation theorem and the heat quantum metric}\label{sect_linear_drive_real}

We now consider linear drives of the form
\begin{equation}
     \delta\hat{H}_{\alpha\beta}^{\pm,\mu\nu}(t)=2 \left(V_{0\alpha}\hat{P}_{\alpha}^{\mu} \pm V_{0\beta}\hat{P}^{\nu}_{\beta}\right) \cos(\omega t).\label{linear_probe_real}
\end{equation}
In this case, the integrated rates naturally reflect the fluctuations of the generalized polarizations in the ground state, through the fluctuation-dissipation theorem~\cite{kubo1957statistical,ozawa2019probing}. In particular, when considering the pure thermal probe ($\alpha,\beta\!=\!Q$)
\begin{equation}
     \delta\hat{H}_{Q}^{\mu}(t)=2 V_{0Q}\hat{P}_{Q}^{\mu}  \cos(\omega t),\label{linear_probe_real_thermal}
\end{equation}
the integrated rates yield the fluctuations of the heat polarization
\begin{equation}
    \frac{\hbar^2}{2\pi V_{0Q}^2}\int_{0}^{\infty}d\omega\,\Gamma_{QQ}^{\mu}(\omega) = {\rm Var} (\hat P_Q^{\mu}) ,\label{fluctuation_dissipation_bis}
\end{equation}
where ${\rm Var} (\cdot)$ denotes the variance in the many-body ground state. Equation~\eqref{fluctuation_dissipation_bis} simply reflects the fluctuation-dissipation theorem applied to the heat polarization ${P}_{Q}^{\mu}$. 

%It is illuminating to connect the result in Eq.~\eqref{fluctuation_dissipation_bis} with the discussion in Section~\ref{sect_linear_drive}, where the integrated rates under linear driving were related to the integrated heat quantum metric [Eq.~\eqref{fluctuation_dissipation}]. Although that earlier result was derived assuming periodic boundary conditions, it continues to hold in open-boundary settings provided the system is in a genuinely insulating phase, i.e., one without metallic edge modes. In this regime, the combined results show that the average heat quantum metric -- the many-body heat quantum metric -- captures the fluctuations of the heat polarization in the ground state

It is illuminating to relate the result in Eq.~\eqref{fluctuation_dissipation_bis} to the discussion in Section~\ref{sect_linear_drive}. There, the integrated rates under linear driving were shown to coincide with the integrated heat quantum metric (or \emph{many-body heat quantum metric}) using periodic boundary conditions; see Eq.~\eqref{fluctuation_dissipation}. One deduces that the many-body heat quantum metric captures the ground-state fluctuations of the heat polarization in open-boundary settings~\footnote{It would be interesting to establish a more general and formal relation between the many-body heat quantum metric and the fluctuations of the heat polarization by generalizing the cumulant-generating-function formalism of Ref.~\cite{souza2000polarization} to the thermal context.}, upon neglecting edge contributions~\cite{tran2017probing,ozawa2018extracting}. 

Besides, we note that the momentum-resolved heat quantum metric $g_{QQ}^{xx}(\bm{k})$ can also be accessed in open‑boundary geometries by applying the probe in Eq.~\eqref{linear_probe_real} to Bloch wave packets and monitoring the resulting excitation rates~\cite{ozawa2018extracting}.

%Following Section~\ref{sect_linear_drive}, the resulting excitation rates would give access to the real part of the Kubo correlators $\mathbb{L}_{\alpha\beta}^{\mathrm{abs},\mu\nu}(\omega)$ within an open boundary setting. 

%- wave-packets: access k-space metric within a realistic geometry

%{\pink The components of the heat quantum metric [Eq.~\eqref{eq_heat_metric}] can be extracted by applying a similar driving scheme, namely, by replacing the chiral drives in Eq.~\eqref{general_probe} by the following linear drives
%\begin{align}
%&g_{QQ}^{xx} \longrightarrow \delta\hat{H}_{Q}^{x}(t) = 2 V_{0 Q}\cos(\omega t)\hat{P}_{Q}^{x}, \notag \\
%&g_{QQ}^{yy} \longrightarrow \delta\hat{H}_{Q}^{y}(t) = 2 V_{0 Q}\cos(\omega t)\hat{P}_{Q}^{y}, \notag \\
%&g_{QQ}^{xy} \longrightarrow \delta\hat{H}_{Q}^{\pm,xy}(t) = 2 V_{0Q}\cos(\omega t)\left(\hat{P}_{Q}^{x} \pm \hat{P}^{y}_{Q}\right),
%\end{align}
%and extracting the corresponding integrated rates. This scheme thus generalizes the quantum-metric-measurement method of Refs.~\cite{ozawa2018extracting,ozawa2019probing,yu2020experimental} to the thermal context.
%}

\subsection{Implementing thermoelectric probes in engineered lattice systems}

We now discuss how the generalized thermoelectric probes in Eqs.~\eqref{general_probe} and \eqref{linear_probe_real} can be implemented in engineered lattice systems, described by a tight-binding Hamiltonian:
\begin{equation}
   \hat{ \mathcal{H}}=\sum_{ij}t_{ij}\hat{c}^\dagger_i\hat{c}_j + U_j \hat{c}^\dagger_j\hat{c}_j \,,
\end{equation}
where $\hat{c}^\dagger_j$ creates a particle at lattice site $j$, where $t_{ij}$ denotes the hopping matrix element connecting lattice sites $i$ and $j$, and where $U_j$ denotes an onsite potential. The corresponding electric polarization operator takes the form of a linear gradient potential,
\begin{equation}
   \hat{P}^\mu_e=-e\sum_{i}r^\mu_i\hat{c}^\dagger_i\hat{c}_i\,,
\end{equation}
while the heat polarization introduced in Eq.~\eqref{pol_real} takes the form of a linear ``gravitational" potential, or strain,
\begin{equation}
   \hat{P}^\mu_Q=\sum_{i,j}\frac{r^\mu_i+r^\mu_j}{2}t_{ij}\hat{c}^\dagger_i\hat{c}_j \,   + \sum_j (U_j - \mu) r_j^{\mu} \hat{c}^\dagger_j\hat{c}_j \,.\label{polar_strain}
\end{equation}
Hence, the perturbations $\delta\hat{H}_{e Q}^{\pm}(t)$ and $\delta\hat{H}_{Q Q}^{\pm}(t)$ in Eqs.~\eqref{general_probe} and \eqref{linear_probe_real} involve a spatially-dependent time-modulation of the bare hopping amplitudes of the form 
%\begin{equation}
%t_{ij}\to t_{ij} \left (1+ 2 V_{0 Q} \left (\frac{r^\mu_i+r^\mu_j}{2}\right ) \cos (\omega t) \right). 
%\end{equation}
\begin{equation}
t_{ij}\to t_{ij} \left (1+ V_{0 Q} \left (r^\mu_i+r^\mu_j\right ) \cos (\omega t) \right), 
\end{equation}
where the strength of the drive should satisfy $(1/V_{0 Q})\!\gg \sqrt{A}$, in order to monitor excitation rates within the perturbative regime. Such a local and time-dependent control over tunneling matrix elements can be achieved in ultracold atomic gases trapped in optical lattices under a quantum-gas microscope; see Refs.~\cite{impertro2024local,nixon2024individually,impertro2025strongly}. The excitation rates [Eqs.~\eqref{Gamma_pm_OBC} and \eqref{fluctuation_dissipation_bis}] could then be measured by monitoring the dynamical repopulation of the Bloch bands through band-mapping, as was previously demonstrated in the dichroic measurement of Ref.~\cite{asteria2019measuring}. We anticipate that similar driving schemes could also be implemented in circuit QED platforms, where high control over tunneling matrix elements can be achieved~\cite{roushan2017chiral,wang2024realization}.  In solid-state materials, time-modulated strain could be activated by exciting phononic modes; see the recent Ref.~\cite{li2025phonondichroismsrevealingunusual} on phonon dichroisms and their relation to heat magnetic moments. 

We remark that the perturbations $\delta\hat{H}_{e Q}^{\pm}(t)$ and $\delta\hat{H}_{Q Q}^{\pm}(t)$ defined in Eqs.~\eqref{general_probe} and \eqref{linear_probe_real} also potentially involve a time-modulated gradient, through the onsite potential entering the heat polarization $\hat{P}^\mu_Q$; see Eq.~\eqref{polar_strain}.

\subsection{Measuring individual contributions to the orbital magnetization}

As previously reminded in Eq.~\eqref{two_contributions}, the total orbital magnetization can be partitioned in terms of two contributions, $\bm{\mathcal{M}}\!=\!\bm{\mathcal{M}}^{{\rm SR}}+\bm{\mathcal{M}}^{{\rm COM}}$, which can be individually addressed by combining two sum rules; see Section~\ref{section_sum_rules}. In practice, one would first extract the total orbital magnetization $\bm{\mathcal{M}}$ through the sum rule in Eq.~\eqref{magn_real}, i.e. by subjecting the system to the chiral probe $\delta\hat{H}_{e Q}^{\pm,\mu\nu}(t)$ and monitoring the corresponding dichroic response. Then, the self-rotation contribution $\mathcal{M}^{{\rm SR}}$ would be extracted by measuring the dichroic response resulting from the conventional chiral probe $\delta\hat{H}_{e e}^{\pm}(t)$, and using the relation
\begin{equation}
  \frac{1}{A(V_{0e})^2}  \int_{0}^{\infty}\! \, \omega \, \Delta\Gamma^{\mu\nu}_{ee}(\omega) d\omega= - \epsilon_{\mu\nu\rho} \frac{4\pi e c}{\hbar^2} \mathcal{M}_\rho^{{\rm SR}},\label{f_sum_rule_bis}
\end{equation}
which directly derives from the $f$-sum rule in Eq.~\eqref{f_sum_rule} and Eq.~\eqref{ImLabs_rates}. The center-of-mass contribution would then be obtained through the difference $\bm{\mathcal{M}}^{{\rm COM}}\!=\!\bm{\mathcal{M}}-\bm{\mathcal{M}}^{{\rm SR}}$.\\

%{\pink [NG: Indicate here how combining different drives and sum rules allows one to isolate the different contributions to the magnetization ?]}

\section{Generalized heat magnetizations and their sum-rule partitions}\label{section_generalization}

In Sections~\ref{section_sum_rules}-\ref{Section_generalized_dichroism}, we established that the self-rotation contribution to the orbital magnetization can be derived from the conventional dichroic $f$-sum rule, whereas the total orbital magnetization is governed by a distinct sum rule involving a thermoelectric probe. In this Section, we extend this sum-rule partitioning to the heat magnetization and its higher-order generalizations. In particular, we reveal a hierarchical structure of sum rules that generalizes the physical partitioning of the orbital magnetization to heat magnetizations.

\subsection{Partitioning the heat magnetization}

Let us start by partitioning the total heat magnetization as 
\begin{equation}
    \bm{\mathcal{M}}^Q=\bm{\mathcal{M}}^{Q,1}+\bm{\mathcal{M}}^{Q,2}+\bm{\mathcal{M}}^{Q,3}\,,\label{heat_partition}
\end{equation}
where
\begin{subequations}
\label{eq:partition}
    \begin{align}
        \mathcal{M}_\rho^{Q,1}&=\!-\frac{i\epsilon_{\rho\mu\nu}}{8\hbar}
\!\!\!\int\!\!\!\frac{d^2k}{(2\pi)^2}\!\!\!\!\sum_{n\in\mathrm{occ}}
\!\langle\partial_{k_\mu}u_{n\bm{k}}|
\left(\!\hat{H}_{\bm{k}}\!-\!\varepsilon_{n\bm{k}}\!\right)^{2}
\!\!|\partial_{k_\nu}u_{n\bm{k}}\rangle \\
        \mathcal{M}_\rho^{Q,2}&=\!\frac{i\epsilon_{\rho\mu\nu}}{2\hbar}
\!\!\!\int\!\!\!\frac{d^2k}{(2\pi)^2}\!\!\!\!\sum_{n\in\mathrm{occ}}
\!\langle\partial_{k_\mu}u_{n\bm{k}}|
\left(\!\!\frac{\hat{H}_{\bm{k}}\!+\!\varepsilon_{n\bm{k}}\!-\!2\mu}{2}\!\!\right)
\!\!\!\\&\quad\quad\quad\quad\quad\quad\quad\quad\times\left(\!\hat{H}_{\bm{k}}\!-\!\varepsilon_{n\bm{k}}\!\right)|\partial_{k_\nu}u_{n\bm{k}}\rangle \nonumber\\
\label{Mq3:com}
        \mathcal{M}_\rho^{Q,3}&=\!\frac{1}{2\hbar}
\int \frac{d^2k}{(2\pi)^2} \sum_{n\in\mathrm{occ}} \left(\varepsilon_{n\bm{k}}-\mu\right)^2\Omega_{n\bm{k}}^\rho\,.
    \end{align}
\end{subequations}

First, we remind that the total heat magnetization $\bm{\mathcal{M}}^Q$ can be expressed in terms of the sum rule in Eq.~\eqref{eq_dichroism_heat}, which involves the heat probe $\delta\hat{H}_{Q Q}^{\pm}(t)$.

Then, we note that the first contribution $\bm{\mathcal{M}}^{Q,1}$ entering Eq.~\eqref{heat_partition} can be expressed in terms of a sum rule associated with the conventional electric probe $\delta\hat{H}^{\pm}_{ee}(t)$, 
\begin{equation}
    \frac{1}{A(V_{0e})^2}  \int_{0}^{\infty}\! \, \omega^2 \, \Delta\Gamma^{\mu\nu}_{ee}(\omega) d\omega=-\epsilon_{\rho\mu\nu}\frac{16\pi e^2}{\hbar^3}\mathcal{M}^{Q,1}_\rho\,.\label{kubo_dichroism_magnetic}
\end{equation}
In fact, this sum rule was originally introduced by Kubo~\cite{kubo1957statistical}:~considering free particles of mass $m$ subjected to a background magnetic field $B_z$, this sum rule was expressed in terms of the electric conductivity tensor as
\begin{equation}
\frac{2}{\pi}\int_0^\infty d\omega\, \omega {\rm Im}\left [\sigma^{xy}(\omega) \right ] = \frac{ne^3}{m^2c}B_z , 
\end{equation}
where $n$ denotes the particle density. A comparison with Eq.~\eqref{kubo_dichroism_magnetic} is obtained by noticing that ${\rm Im}\left[\sigma^{xy}(\omega)\right] \sim \omega \Delta \Gamma^{xy}_{ee}(\omega)$; see Eq.~\eqref{eq:power_absorption}.
 We point out that similar sum rules were derived in the context of rotating Bose-Einstein condensates~\cite{zambelli1998quantized,pitaevskii2016bose}, relating the angular momentum of a rotating atomic gas to the dynamical structure factor relative to quadrupolar excitations.

The second contribution $\bm{\mathcal{M}}^{Q,2}$ accounts for the intrinsic ``heat orbital moment" (or \emph{heat-self-rotation}) contribution to the heat magnetization. Indeed, this semiclassical interpretation of $\bm{\mathcal{M}}^{Q,2}$ was revealed in Ref.~\cite{zhang2020thermodynamics}, by evaluating the expectation value of the heat-orbital-moment operator,
\begin{equation}
\hat{\bm{m}}^Q =\frac{1}{4} (\hat{\bm{r}} - {\bm r}_c) \times \hat{\bm{j}}_Q (\hat{\bm{r}}) + {\rm h.c.},
\end{equation}
in a Bloch-state wave packet, where $\hat{\bm{j}}_Q$ denotes the local heat current density. We remark that this quantity generalizes the orbital magnetic moment of Bloch states~\cite{xiao2010berry} to the thermal context, by simply replacing the electric current operator with the heat current operator.

In direct analogy with the dichroic $f$-sum rule in Eqs.~\eqref{f_sum_rule} and \eqref{f_sum_rule_bis}, the contribution $\bm{\mathcal{M}}^{Q,2}$ can be extracted by measuring the dichroic response resulting from the chiral thermoelectric probe $\delta\hat{H}^{\pm}_{eQ}(t)$, through the relation 
\begin{equation}
    \frac{1}{AV_{0e}V_{0Q}}  \int_{0}^{\infty}\! \, \omega \, \Delta\Gamma^{\mu\nu}_{eQ}(\omega) d\omega=-\epsilon_{\rho\mu\nu}\frac{4\pi e}{\hbar^2}\mathcal{M}^{Q,2}_\rho\,.\label{heat_self}
\end{equation}
Interestingly, from Eq.~\eqref{eq_dichroism_orbital}, we notice that the integrand on the LHS of Eq.~\eqref{heat_self} can be interpreted as the frequency-resolved orbital magnetization, $\mathcal{M}_\rho (\omega)$, formally defined through
\begin{align}
&\int_{0}^{\infty}\! \,\mathcal{M}_\rho (\omega) d\omega = \mathcal{M}_\rho, \notag \\
&\mathcal{M}_\rho (\omega) \equiv \frac{1}{AV_{0e}V_{0Q}}\frac{\hbar}{4\pi c}  \epsilon_{\mu\nu\rho}\, \Delta\Gamma^{\mu\nu}_{eQ}(\omega).\label{freq_res_magnet} 
\end{align}
Hence, as far as the self-rotation contribution $\mathcal{M}^{Q,2}$ is concerned, the \emph{heat} magnetization can be understood as resulting from a sum rule involving a frequency integral of the \emph{orbital} magnetization
\begin{equation}
     \int_{0}^{\infty}\! \, \omega \mathcal{M}_\rho (\omega) d\omega=-\frac{2\pi}{\Phi_0}\mathcal{M}^{Q,2}_\rho\,,\label{heat_self_bis}
\end{equation}
where we combined Eqs.~\eqref{heat_self}-\eqref{freq_res_magnet}.

Altogether, in analogy with the center-of-mass contribution to the orbital magnetization in Eq.~\eqref{orb_COM}, the third contribution $\bm{\mathcal{M}}^{Q,3}$ constitutes the ``new" piece of information, not directly accessible through a dedicated sum rule. This ``center-of-mass" contribution can nevertheless be evaluated through the three sum rules listed above, combined with the relation $\bm{\mathcal{M}}^{Q,3}\!=\!\bm{\mathcal{M}}^{Q}\!-\!\bm{\mathcal{M}}^{Q,1}\!-\!\bm{\mathcal{M}}^{Q,2}$.

\subsection{General hierarchical construction}
The partitioning of the orbital and heat magnetizations in terms of sum rules and ``center-of-mass" contributions can be systematically generalized to a whole hierarchy of higher-order heat magnetizations. In the present construction, the ``order" refers to the power with which heat enters the generalized magnetizations, as we now describe.

Our construction starts by introducing the generalized polarization operators
\begin{equation}
    \hat{P}^\mu_n=\frac{1}{2^n}\sum_{k=0}^n\begin{pmatrix}
        n\\k
    \end{pmatrix}\left(\hat{H}-\mu\right)^k\hat{r}^\mu\left(\hat{H}-\mu\right)^{n-k},\label{gen_pol_eq}
\end{equation}
where one recognizes the conventional electric polarization $\hat{P}^\mu_e\!=\!-e\hat{P}^\mu_0$ and heat polarization $\hat{P}^\mu_Q\!=\!\hat{P}^\mu_1$.

Then, inspired by the Kramers-Kronig relations in Eq.~\eqref{L_DC_T0_final}, one defines the $n$th-order heat magnetization density as
\begin{equation}
\label{eq:hier}
  -\,\epsilon^{\mu\nu\rho}\mathcal{M}^{(n)}_{\rho} \equiv   \mathbb{L}_{\alpha\beta}^{\mathrm{DC},\mu\nu}, \quad n=\alpha+\beta ,
\end{equation}
 where the correlation functions $\mathbb{L}_{\alpha\beta}^{\mathrm{DC}}$ [Eq.~\eqref{inter_Lee}] now include the generalized polarization operators $\hat{P}^\mu_{\alpha}$ and $\hat{P}^\mu_{\beta}$ defined in Eq.~\eqref{gen_pol_eq}.

 In Bloch representation, these quantities are explicitly given by the $\bm{k}$-space integral 
\begin{equation}
\label{eq:hier_k}
    \mathcal{M}^{(m)}_\rho=\!\frac{i\epsilon_{\rho\mu\nu}}{\hbar}
\!\!\!\int\!\!\!\frac{d^2k}{(2\pi)^2}\!\!\!\!\sum_{n\in\mathrm{occ}}
\!\langle\partial_{k_\mu}u_{n\bm{k}}|
\left(\!\!\frac{\hat{H}_{\bm{k}}\!+\!\varepsilon_{n\bm{k}}\!-\!2\mu}{2}\!\!\right)^{m}
\!\!\!|\partial_{k_\nu}u_{n\bm{k}}\rangle\,.
\end{equation}
By considering the first orders $m\!=\!0,1,2$, one recognizes the expressions for $C$, $\mathcal{M}$ and $\mathcal{M}^Q$, namely, $\mathcal{C}_\rho=\frac{2\pi}{\hbar}\mathcal{M}^{(0)}_\rho$, $\mathcal{M}_\rho=\frac{-e}{c}\mathcal{M}^{(1)}_\rho$ and $\mathcal{M}^Q_\rho=\frac{1}{2}\mathcal{M}^{(2)}_\rho$. 

We then generalize the chiral probe $\delta\hat{H}_{\alpha\beta}^{\pm}(t)$ in Eq.~\eqref{general_probe} so as to encompass the generalized polarization operators, $\hat{P}^\mu_{\alpha}$ and $\hat{P}^\mu_{\beta}$, with $\alpha,\beta\!\in\!\mathbb{N}$. We find that the corresponding dichroic responses yield the higher-order heat magnetizations $\mathcal{M}^{(\alpha + \beta)}$ through the relation
\begin{equation}
    \begin{aligned}
        \frac{\Delta\Gamma^{int,\mu\nu}_{\alpha\beta}}{A}&=\int_0^\infty d\omega\,\frac{\Delta\Gamma^{\mu\nu}_{\alpha\beta}(\omega)}{A}\\&=\frac{2\pi}{\hbar}V_{0\alpha}V_{0\beta}\epsilon_{\mu\nu\rho}\mathcal{M}^{(\alpha+\beta)}_\rho .\label{gen_pol_eq_bis}
    \end{aligned}
\end{equation}

Importantly, similarly to the orbital and heat magnetizations discussed above, the information carried by these higher-order heat magnetizations is redundant in the sense that most of their contributions can be inferred from sum rules involving (frequency-resolved) lower-order magnetizations. To see this, let us introduce the quantity
\begin{equation}
    \begin{aligned}
m_\rho^{a,b}&=\!\frac{i\epsilon_{\rho\mu\nu}}{\hbar}
\!\!\!\int\!\!\!\frac{d^2k}{(2\pi)^2}\!\!\!\!\sum_{n\in\mathrm{occ}}
\!\langle\partial_{k_\mu}u_{n\bm{k}}|
\left(\!\!\frac{\hat{H}_{\bm{k}}\!+\!\varepsilon_{n\bm{k}}\!-\!2\mu}{2}\!\!\right)^a
\!\!\!\\&\quad\quad\quad\quad\quad\quad\quad\quad\times\left(\!\hat{H}_{\bm{k}}\!-\!\varepsilon_{n\bm{k}}\!\right)^b|\partial_{k_\nu}u_{n\bm{k}}\rangle .
    \end{aligned}
\end{equation}
One then identifies an insightful partitioning of the $n$th-order heat magnetization expressed as
\begin{equation}
    \mathcal{M}_\rho^{(n)}=m_\rho^{n,0}=\mathcal{M}^{{\rm COM},(n)}_\rho-\sum_{k=1}^n\begin{pmatrix}
        n\\k
    \end{pmatrix}\left(\frac{-1}{2}\right)^{\!\!k}m_\rho^{n-k,k}.\label{general_partition}
\end{equation}
Importantly, all the contributions entering Eq.~\eqref{general_partition} can be expressed as generalized sum rules  involving lower-order (frequency-resolved) dichroic responses,
\begin{equation}
    \begin{aligned}
m_\rho^{a,b}&=\epsilon_{\rho\mu\nu}\int_0^\infty d\omega\,\left(\hbar\omega\right)^b\frac{\Delta\Gamma^{\mu\nu}_{\alpha\beta}(\omega)}{V_{0\alpha}V_{0\beta}}, \quad \alpha+\beta=a ,
    \end{aligned}\label{eq_freq_res_gen}
\end{equation}
except for the ``center-of-mass" contribution 
\begin{equation}
    \mathcal{M}_\rho^{{\rm COM},(m)}=\frac{1}{\hbar}\int \frac{d^2k}{(2\pi)^2} \sum_{n\in\mathrm{occ}} \left(\varepsilon_{n\bm{k}}-\mu\right)^m\Omega_{n\bm{k}}^\rho ,
\end{equation}
which contains the ``new" information at the $n$-th order, not captured by a dedicated sum rule involving lower-order quantities. 

We stress that Eq.~\eqref{eq_freq_res_gen} generalizes the sum rules for the individual contributions $\bm{\mathcal{M}}^{Q,1}$ and $\bm{\mathcal{M}}^{Q,2}$ to the heat magnetization in Eqs.~\eqref{kubo_dichroism_magnetic} and \eqref{heat_self}, as well as the dichroic $f$-sum rule for the self-rotation contribution $\bm{\mathcal{M}}^{{\rm SR}}$ to the orbital magnetization in Eq.~\eqref{f_sum_rule_bis}. The general partitioning expressed in Eq.~\eqref{general_partition} thus emerges as a foundational principle governing the sum rules established in this work.

\section{Concluding remarks}\label{section_conclusion}

This work presented a unified framework connecting the orbital and heat magnetizations to experimentally accessible chiral probes via generalized sum rules, placing these fundamental ground-state properties on the same footing as the topological Chern number.

We have set a particular emphasis on the dichroic response to chiral probes $\delta\hat{H}_{\alpha\beta}^{\pm}(t)$, as defined in Eq.~\eqref{general_probe}, which naturally lead to the extraction of the total orbital and heat magnetizations. As we discussed, this framework represents a significant extension to the dichroic $f$-sum rule and to similar dichroic responses leading to the Berry curvature and Chern number of Bloch band systems~\cite{souza2008dichroic,tran2017probing,repellin2019detecting,asteria2019measuring,goldman2024relating}. However, more general probes and sum rules could be considered within this framework. Indeed, the Kramers–Kronig relations in Eq.~\eqref{L_DC_T0_final} can be extended by considering a broader class of integrals of the form
\begin{align}
\int_{0}^{\infty}\!d\omega\frac{\mathrm{Im}\left[\mathbb{L}^{\mathrm{abs},\mu\nu}_{\alpha\beta}(\omega)\right]}{\omega^n} , \quad
\int_{0}^{\infty}\!d\omega\frac{\mathrm{Re}\left[\mathbb{L}^{\mathrm{abs},\mu\nu}_{\alpha\beta}(\omega)\right]}{\omega^n}, 
\end{align}
where $n$ is an integer, and where the (absorptive) correlation functions are defined in Eq.~\eqref{Labs}. Setting $n\!=\!1$ naturally leads to generalizations of the quantum geometric tensor~\cite{kolodrubetz2017geometry,shinada2025quantum}, with the real and imaginary parts giving access to possible analogues of the Berry curvature and quantum metric, respectively. As for the conventional quantum geometric tensor, the real and imaginary parts would be extracted by monitoring excitation rates under linear and circular drives, respectively~\cite{ozawa2018extracting,ozawa2019probing,yu2020experimental}. 
%For instance, measuring the excitation rates $\Gamma_Q^x (\omega)$ resulting from the linear drive $\delta\hat{H}(t)\!=\! 2 V_{0Q} \hat{P}_Q^x \cos (\omega t)$, would inform on the fluctuations of the heat polarization through the fluctuation-dissipation theorem
%\begin{equation}
%\int_0^{\infty} \Gamma_Q^x (\omega) d\omega = \frac{2 \pi (V_{0Q})^2}{\hbar^2} \, {\rm Var}  (\hat P_Q),\label{heat_diss_fluct}
%\end{equation}
%where ${\rm Var}(\hat P_Q)\!=\! \langle \hat P_Q ^2 \rangle - \langle \hat P_Q \rangle^2$ denotes the variance of the operator. The protocol leading to Eq.~\eqref{heat_diss_fluct} would yield the diagonal component $g_{Q}^{xx}\!\equiv\!{\rm Var} (\hat P_Q)$ of the ``heat quantum metric" $g_{Q}^{\mu \nu}$. The other components of this ``heat metric" could be extracted by varying the orientation of the linear drive~\cite{ozawa2018extracting,ozawa2019probing}. 
Establishing generalized quantum geometric tensors -- for instance, to identify geometric bounds of ground-state properties~\cite{shinada2025quantum} -- and further elucidating the properties of the “heat quantum metric” introduced in Eq.~\eqref{eq_heat_metric} represent compelling directions for future investigation; see the recent works~\cite{Buthenhoff2026,lhachemi2026unifyingframeworksumrules}.

Our study has primarily examined insulating phases at zero temperature in the absence of interactions. A natural extension of the framework would be its application to correlated insulators. While the sum rules derived in this work [e.g., Eqs.~\eqref{L_DC_T0_final} and \eqref{DIR2}] are expected to remain valid in the presence of interactions, the corresponding heat-current and heat-polarization operators would necessarily incorporate these processes explicitly~\cite{Cooper1997}. Moreover, it would be interesting to elucidate how our momentum-space relations [e.g., Eqs.~\eqref{defs_CM}-\eqref{eq_heat_metric}] would be reformulated using twisted boundary conditions, which provide an appropriate parameter space for many-body settings~\cite{souza2000polarization,ozawa2019probing,goldman2024relating}. 

As noted above, our framework relies on zero-temperature transport coefficients, rendering finite-temperature generalizations an intriguing challenge. 
%We nevertheless expect the results to remain reasonably accurate at temperatures small compared to the insulating gap. 
Within this context, elucidating the geometric content of thermoelectric sum rules at finite temperature -- particularly the fate of the heat quantum metric in Eq.~\eqref{eq_heat_metric} when extended to mixed states -- emerges as a compelling open problem~\cite{ji2025density}.

%We further anticipate that our results should apply to metals at zero temperature, noting that magnetizations remain governed by Kubo correlators in this limit, as expressed in Eq.~\eqref{transp_coeff}.
Moreover, while the physical quantities analyzed here are integrated over the entire Brillouin zone, it would be valuable to examine their momentum-resolved components, accessible either through the dichroic response of localized wave packets~\cite{tran2017probing} or momentum-resolved spectroscopy~\cite{schuler2020local,beaulieu2024berry}. We further note that the applicability of our thermoelectric dichroic approach may also prove relevant in the study of bosonic systems~\cite{bermond2025local,tesfaye2025quantumgeometrybosonicbogoliubov}. 

Another promising direction concerns the study of edge contributions to ground-state properties and thermoelectric dichroic signals, for instance, by making use of effective low-energy edge theories~\cite{unal2024quantized}. Finally, it would be worthwhile to identify concrete dichroic schemes for extracting the orbital magnetization density of Floquet systems, where this quantity is quantized according to Floquet winding numbers~\cite{Nathan2017,Gavensky2025,gavensky2025quantizedchernsimonsaxioncoupling}.\\

%\BB{Shall we add a comment on the possibility to use these protocole to probe the hall viscosity via the streda formula \cite{hidaka2013viscoelastic} ?}

\begin{acknowledgments}
%\textbf{\textit{Acknowledgments.---}} 
We thank C\'ecile Repellin and BoYe Sun for discussions at the early stage of the project. We also thank Nigel Cooper for his insightful comments on our manuscript. This research was financially supported by the ERC Grant LATIS, the EOS project CHEQS and the Fondation ULB. This work also received support from the French government, managed by the National Research Agency, under the France 2030 program, reference ANR‑23‑PETQ‑0002.L.P.G. acknowledges support provided by the FRS-FNRS Belgium and the L'Or\'eal-UNESCO for Women in Science Programme.
\end{acknowledgments}


%\bibliography{mibib}

%%%%%%%%%%%%%%%
\appendix
%\section{...Eq28?}

\begin{widetext}
\section{Partitioning the heat magnetization}
In this appendix, we show how to derive the partitioning of the heat magnetization presented in Eq.~\eqref{eq:partition}. We discuss its physical interpretation and the related sum-rules in more detail.\\

We start with the expression for the heat magnetization in Eq.~\eqref{MQ_def},
\begin{equation}
    \mathcal{M}^{Q}_{\rho}
=\!\frac{i\epsilon_{\rho\mu\nu}}{2\hbar}
\!\!\!\int\!\!\!\frac{d^2k}{(2\pi)^2}\!\!\!\!\sum_{n\in\mathrm{occ}}
\!\langle\partial_{k_\mu}u_{n\bm{k}}|
\left(\!\!\frac{\hat{H}_{\bm{k}}\!+\!\varepsilon_{n\bm{k}}\!-\!2\mu}{2}\!\!\right)^{2}
\!\!\!|\partial_{k_\nu}u_{n\bm{k}}\rangle\,.
\end{equation}
Then, inspired by the partition of the orbital magnetization in terms of a self-rotating and center-of-mass contribution, we decompose the heat magnetization by writing 
\begin{equation}
    \left(\frac{\hat{H}_{\bm{k}}+\varepsilon_{n\bm{k}}-2\mu}{2}\right)^2=-\frac{1}{4}\left(\hat{H}_{\bm{k}}-\varepsilon_{n\bm{k}}\right)^2+\left(\frac{\hat{H}_{\bm{k}}+\varepsilon_{n\bm{k}}-2\mu}{2}\right)\left(\hat{H}_{\bm{k}}-\varepsilon_{n\bm{k}}\right)+\left(\varepsilon_{n\bm{k}}-\mu\right)^2
\end{equation}
such that,
\begin{equation}
    \bm{\mathcal{M}}^Q=\bm{\mathcal{M}}^{Q,1}+\bm{\mathcal{M}}^{Q,2}+\bm{\mathcal{M}}^{Q,3}\,,
\end{equation}
with
\begin{subequations}
    \begin{align}
        \mathcal{M}_\rho^{Q,1}&=-\frac{i\epsilon_{\rho\mu\nu}}{8\hbar}\int\frac{d^2k}{(2\pi)^2}\sum_{n\in\mathrm{occ}}\langle\partial_{k_\mu}u_{n\bm{k}}|\left(\hat{H}_{\bm{k}}-\varepsilon_{n\bm{k}}\right)^{2}|\partial_{k_\nu}u_{n\bm{k}}\rangle \\
        \mathcal{M}_\rho^{Q,2}&=\frac{i\epsilon_{\rho\mu\nu}}{2\hbar}\int\frac{d^2k}{(2\pi)^2}\sum_{n\in\mathrm{occ}}\langle\partial_{k_\mu}u_{n\bm{k}}|\left(\frac{\hat{H}_{\bm{k}}+\varepsilon_{n\bm{k}}-2\mu}{2}\right)\left(\!\hat{H}_{\bm{k}}-\varepsilon_{n\bm{k}}\right)|\partial_{k_\nu}u_{n\bm{k}}\rangle \nonumber\\
        \mathcal{M}_\rho^{Q,3}&=\frac{i\epsilon_{\rho\mu\nu}}{2\hbar}\int \frac{d^2k}{(2\pi)^2} \sum_{n\in\mathrm{occ}} \left(\varepsilon_{n\bm{k}}-\mu\right)^2\langle\partial_{k_\mu}u_{n\bm{k}}|\partial_{k_\nu}u_{n\bm{k}}\rangle\\&=\frac{1}{2\hbar}\int \frac{d^2k}{(2\pi)^2} \sum_{n\in\mathrm{occ}} \left(\varepsilon_{n\bm{k}}-\mu\right)^2\Omega_{n\bm{k}}^\rho\,.
    \end{align}
\end{subequations}
By comparing these terms with those obtained in the partitioning of the heat magnetization together with the differential rates, it is then possible to provide an interpretation for these terms:
\begin{itemize}
    \item Starting with $\bm{\mathcal{M}}^{Q,1}$, we note that upon a frequency integration and a resolution of the identity
    \begin{equation}
        \begin{aligned}
            \frac{1}{A(V_{0e})^2}\int_0^{+\infty}\omega^2 \, \Delta\Gamma^{\mu\nu}_{ee}(\omega) d\omega&=-\frac{2\pi}{\hbar^2}\int\frac{d^2k}{(2\pi)^2}\int_0^{+\infty}d\omega\,\omega^2\sum_{n\in\textrm{occ}}\sum_{m\in\textrm{unocc}}\,P^{-,\mu\nu}_{ee,nm}(\bm{k})\delta(\omega-\omega_{mn}(\bm{k})) \\
            &=-\frac{4\pi}{\hbar^2}\int\frac{d^2k}{(2\pi)^2}\sum_{n\in\textrm{occ}}\sum_{m\in\textrm{unocc}}\,\omega_{mn}^2(\bm{k})\mathrm{Im}\left[\langle u_{n\bm{k}}|\hat{P}^{\mu}_{e}|u_{m\bm{k}}\rangle \langle u_{m\bm{k}}|\hat{P}^{\nu}_{e}|u_{n\bm{k}}\rangle\right]\\
            &=-\frac{4\pi e^2}{\hbar^4}\int\frac{d^2k}{(2\pi)^2}\sum_{n\in\textrm{occ}}\sum_{m\in\textrm{unocc}}\,\mathrm{Im}\left[\langle \partial_{k^\mu}u_{n\bm{k}}|\left(\hat{H}_{\bm{k}}-\varepsilon_{n\bm{k}}\right)^2|u_{m\bm{k}}\rangle \langle u_{m\bm{k}}|\partial_{k^\nu}u_{n\bm{k}}\rangle\right]\\
            &=-\frac{4\pi e^2}{\hbar^4}\int\frac{d^2k}{(2\pi)^2}\sum_{n\in\textrm{occ}}\,\mathrm{Im}\left[\langle \partial_{k^\mu}u_{n\bm{k}}|\left(\hat{H}_{\bm{k}}-\varepsilon_{n\bm{k}}\right)^2|\partial_{k^\nu}u_{n\bm{k}}\rangle\right]\\
            &=-\epsilon_{\rho\mu\nu}\frac{16\pi e^2}{\hbar^3}\mathcal{M}^{Q,1}_\rho\,.
        \end{aligned}
    \end{equation}
    Upon rewriting $\mathrm{Im}(\sigma^{\mu\nu}(\omega))=-\hbar\omega\frac{\Delta\Gamma^{\mu\nu}_{ee}}{A(V_{0e})^2}$ (see Eq.~\eqref{eq:power_absorption}), we recover the interpretation of $\bm{\mathcal{M}}^{Q,1}$ in terms of an electric conductivity tensor sum-rule
    \begin{equation}
        \mathcal{M}_{\rho}^{Q,1}=\frac{\hbar^2}{8\pi e^2}\epsilon_{\rho\mu\nu}\int_0^{+\infty}\omega\,\sigma^{\mu\nu}(\omega)\,d\omega\,.
    \end{equation}
    Another possible interpretation of this component as a marker of the non-commutativity of the underlying system can be obtained by noticing that
    \begin{equation}
        \begin{aligned}
            \textrm{Tr}\left(-\frac{i}{8\hbar}\epsilon_{\lambda\mu\nu}\hat{v}^\mu\hat{v}^{\nu}.\rho\right)
            &=-\frac{i}{4\hbar}\epsilon_{\lambda\mu\nu}\int\frac{d^2\bm{k}}{(2\pi)^2}\sum_{n\in\textrm{occ}}\sum_{m\in\textrm{unocc}}\langle u_{n\bm{k}}|\hat{v}^\mu|u_{m\bm{k}}\rangle\langle u_{m\bm{k}}|\hat{v}^{\nu}|u_{n\bm{k}}\rangle\\
            &=-\frac{i}{8\hbar}\epsilon_{\lambda\mu\nu}\int\frac{d^2\bm{k}}{(2\pi)^2}\sum_{n\in\textrm{occ}}\sum_{m\in\textrm{unocc}}\hbar^2\omega_{mn}^2(\bm{k})\langle u_{n\bm{k}}|\hat{x}^\mu|u_{m\bm{k}}\rangle\langle u_{m\bm{k}}|\hat{x}^{\nu}|u_{n\bm{k}}\rangle\\
            &=\mathcal{M}_\lambda^{Q,1}
        \end{aligned}
    \end{equation}
    \item Similarly, for $\mathcal{M}_\rho^{Q,2}$, we note that upon a frequency integration and a resolution of the identity
    \begin{equation}
        \begin{aligned}
            \frac{1}{AV_{0e}V_{0Q}}\int_0^{+\infty}\omega \, \Delta\Gamma^{\mu\nu}_{Qe}(\omega) d\omega&=-\frac{2\pi}{\hbar^2}\int\frac{d^2k}{(2\pi)^2}\int_0^{+\infty}d\omega\,\omega\sum_{n\in\textrm{occ}}\sum_{m\in\textrm{unocc}}\,P^{-,\mu\nu}_{Qe,nm}(\bm{k})\delta(\omega-\omega_{mn}(\bm{k})) \\
            &=-\frac{4\pi}{\hbar^2}\int\frac{d^2k}{(2\pi)^2}\sum_{n\in\textrm{occ}}\sum_{m\in\textrm{unocc}}\,\omega_{mn}(\bm{k})\mathrm{Im}\left[\langle u_{n\bm{k}}|\hat{P}^{\mu}_{Q}|u_{m\bm{k}}\rangle \langle u_{m\bm{k}}|\hat{P}^{\nu}_{e}|u_{n\bm{k}}\rangle\right]\\
            &=\frac{4\pi e}{\hbar^3}\int\frac{d^2k}{(2\pi)^2}\sum_{n\in\textrm{occ}}\sum_{m\in\textrm{unocc}}\,\mathrm{Im}\left[\langle \partial_{k^\mu}u_{n\bm{k}}|\left(\hat{H}_{\bm{k}}-\varepsilon_{n\bm{k}}\right)\left(\!\frac{\hat{H}_{\bm{k}}+\varepsilon_{n\bm{k}}-2\mu}{2}\!\right)|u_{m\bm{k}}\rangle \langle u_{m\bm{k}}|\partial_{k^\nu}u_{n\bm{k}}\rangle\right]\\
            &=\frac{4\pi e}{\hbar^3}\int\frac{d^2k}{(2\pi)^2}\sum_{n\in\textrm{occ}}\,\mathrm{Im}\left[\langle \partial_{k^\mu}u_{n\bm{k}}|\left(\hat{H}_{\bm{k}}-\varepsilon_{n\bm{k}}\right)\left(\!\frac{\hat{H}_{\bm{k}}+\varepsilon_{n\bm{k}}-2\mu}{2}\!\right)|\partial_{k^\nu}u_{n\bm{k}}\rangle\right]\\
            &=-\epsilon_{\rho\mu\nu}\frac{4\pi e}{\hbar^2}\mathcal{M}_\rho^{Q,2}\,.
        \end{aligned}
    \end{equation}
    Beyond this interpretation of $\mathcal{M}_\rho^{Q,2}$ in terms of a sum-rule, it is also possible to interpret it as a heat-magnetic-moment (or \textit{heat-self-rotation}) in direct analogy with the intrinsic magnetic-moment (or \textit{self rotation}) \cite{zhang2020thermodynamics}. Indeed, defining the heat self-rotation as the cross product of the position of a wave-packet and its center of mass $m^\mu_Q=\frac{1}{4}\epsilon_{\mu\nu\rho}\lbrace\hat{x}^\nu,\hat{J}_Q^\rho\rbrace$, the contribution of the heat self-rotation to the total heat magnetization reads
    \begin{equation}
        \begin{aligned}
            \textrm{Tr}\left(m^\mu_Q.\rho\right)&=\frac{1}{8}\epsilon_{\mu\nu\lambda}\int\frac{d^2\bm{k}}{(2\pi)^2}\sum_{n\in\textrm{occ}}\sum_{m\in\textrm{unocc}}\langle u_{n\bm{k}}|\hat{x}^\nu|u_{m\bm{k}}\rangle\langle u_{m\bm{k}}|\lbrace \hat{H}_{\bm{k}}-\mu,\hat{v}^{\lambda}\rbrace|u_{n\bm{k}}\rangle\\
            &\quad\quad\quad\quad\quad\quad\quad\quad\quad\quad\quad\quad\quad+\langle u_{n\bm{k}}|\lbrace \hat{H}_{\bm{k}}-\mu,\hat{v}^{\lambda}\rbrace|u_{m\bm{k}}\rangle\langle u_{m\bm{k}}|\hat{x}^\nu|u_{n\bm{k}}\rangle\\
            &=\frac{1}{8i\hbar}\epsilon_{\mu\nu\lambda}\int\frac{d^2\bm{k}}{(2\pi)^2}\sum_{n\in\textrm{occ}}\sum_{m\in\textrm{unocc}}\langle u_{n\bm{k}}|\hat{x}^\nu|u_{m\bm{k}}\rangle\langle u_{m\bm{k}}|\left[\hat{x}^{\lambda}, \left(\hat{H}_{\bm{k}}-\mu\right)^2\right]|u_{n\bm{k}}\rangle\\
            &\quad\quad\quad\quad\quad\quad\quad\quad\quad\quad\quad\quad\quad+\langle u_{n\bm{k}}|\left[\hat{x}^{\lambda}, \left(\hat{H}_{\bm{k}}-\mu\right)^2\right]|u_{m\bm{k}}\rangle\langle u_{m\bm{k}}|\hat{x}^\nu|u_{n\bm{k}}\rangle\\
            &=-\frac{i}{4\hbar}\epsilon_{\mu\nu\lambda}\int\frac{d^2\bm{k}}{(2\pi)^2}\sum_{n\in\textrm{occ}}\sum_{m\in\textrm{unocc}}\left((\varepsilon_{n\bm{k}}-\mu)^2-(\varepsilon_{m\bm{k}}-\mu)^2\right)\langle u_{n\bm{k}}|\hat{x}^\nu|u_{m\bm{k}}\rangle\langle u_{m\bm{k}}|\hat{x}^{\lambda}|u_{n\bm{k}}\rangle\\
            &=-\frac{i}{2\hbar}\epsilon_{\mu\nu\lambda}\int\frac{d^2\bm{k}}{(2\pi)^2}\sum_{n\in\textrm{occ}}\sum_{m\in\textrm{unocc}}\left(\varepsilon_{n\bm{k}}-\varepsilon_{m\bm{k}}\right)\left(\frac{\varepsilon_{n\bm{k}}+\varepsilon_{m\bm{k}}-2\mu}{2}\right)\langle \partial_{k^\nu}u_{n\bm{k}}|u_{m\bm{k}}\rangle\langle u_{m\bm{k}}|\partial_{k^\mu}u_{n\bm{k}}\rangle\\
            &=-\frac{i}{2\hbar}\epsilon_{\mu\nu\lambda}\int\frac{d^2\bm{k}}{(2\pi)^2}\sum_{n\in\textrm{occ}}\langle \partial_{k^\nu}u_{n\bm{k}}|\left(\varepsilon_{n\bm{k}}-\hat{H}_{\bm{k}}\right)\left(\frac{\varepsilon_{n\bm{k}}+\hat{H}_{\bm{k}}-2\mu}{2}\right)|\partial_{k^\mu}u_{n\bm{k}}\rangle\\
            &=\mathcal{M}_\lambda^{Q,2}
        \end{aligned}
    \end{equation}
\end{itemize}
\section{A hierarchy of generalized heat magnetization}
Building on the notion of the heat-magnetization, we defined in the main text a hierarchy of generalized heat magnetizations via Eqs.~\eqref{eq:hier}-\eqref{eq:hier_k}, that we partitioned via generalized sum-rules. In this appendix, we will show how to relate the different definitions of these magnetizations [Eqs.~\eqref{eq:hier}-\eqref{eq:hier_k}] and show how to relate them via sum-rules following Eq.~\eqref{general_partition}.
\subsection{Generalized sum-rules and DC Kubo coefficients}

One constructs the generalized magnetization by extending the procedure used for the DC Kubo coefficients -- originally defined through the polarization operators in Eqs.~\eqref{pol_real} -- to the generalized polarization operators of Eq.~\eqref{gen_pol_eq}.
\begin{equation}
    \hat{P}^\mu_n=\frac{1}{2^n}\sum_{k=0}^n\begin{pmatrix}
        n\\k
    \end{pmatrix}\left(\hat{H}-\mu\right)^k\hat{r}^\mu\left(\hat{H}-\mu\right)^{n-k}.
\end{equation}
By following the same steps as between Eq.~\eqref{inter_Lee} to~\eqref{Kramers-Kronig_L} for these generalized polarization, we define
\begin{equation}
  -\,\epsilon^{\mu\nu\rho}\mathcal{M}^{(n)}_{\rho} \equiv   \mathbb{L}_{\alpha\beta}^{\mathrm{DC},\mu\nu}, \quad n=\alpha+\beta .
\end{equation}
More concretely, and considering a k-space representation,
\begin{equation}
  \begin{aligned}
      \mathbb{L}_{\alpha\beta}^{\mathrm{DC},\mu\nu}&=\frac{1}{\hbar}\int\frac{d^2\bm{k}}{(2\pi)^2}\sum_{n \in \textrm{occ}}\sum_{m\in \textrm{unocc}}P^{-,\mu\nu}_{\alpha\beta,nm}(\bm{k})\\
      &=\frac{2}{\hbar}\int\frac{d^2\bm{k}}{(2\pi)^2}\sum_{n \in \textrm{occ}}\sum_{m\in \textrm{unocc}}\textrm{Im}\left[\langle u_{n\bm{k}}|\hat{P}^{\mu}_{\alpha}|u_{m\bm{k}}\rangle \langle u_{m\bm{k}}|\hat{P}^{\nu}_{\beta}|u_{n\bm{k}}\rangle\right]\\
      &=\frac{2}{\hbar}\int\frac{d^2\bm{k}}{(2\pi)^2}\sum_{n \in \textrm{occ}}\sum_{m\in \textrm{unocc}}\sum_{j=0}^\alpha\begin{pmatrix}
        \alpha\\j
    \end{pmatrix}\sum_{l=0}^\beta\begin{pmatrix}
        \beta\\l
    \end{pmatrix}\left(\frac{\varepsilon_{n\bm{k}}-\mu}{2}\right)^{j+l}\left(\frac{\varepsilon_{m\bm{k}}-\mu}{2}\right)^{\alpha+\beta-j-l}\textrm{Im}\left[\langle u_{n\bm{k}}|\hat{r}^{\mu}|u_{m\bm{k}}\rangle \langle u_{m\bm{k}}|\hat{r}^{\nu}|u_{n\bm{k}}\rangle\right]\\
    &=\frac{2}{\hbar}\int\frac{d^2\bm{k}}{(2\pi)^2}\sum_{n \in \textrm{occ}}\sum_{m\in \textrm{unocc}}\left(\frac{\varepsilon_{n\bm{k}}+\varepsilon_{m\bm{k}}-2\mu}{2}\right)^{\alpha+\beta}\textrm{Im}\left[\langle \partial_{k^\mu}u_{n\bm{k}}|u_{m\bm{k}}\rangle \langle u_{m\bm{k}}|\partial_{k^\nu}u_{n\bm{k}}\rangle\right]\\
     &=\frac{2}{\hbar}\int\frac{d^2\bm{k}}{(2\pi)^2}\sum_{n \in \textrm{occ}}\textrm{Im}\left[\langle \partial_{k^\mu}u_{n\bm{k}}|\left(\frac{\hat{H}_{\bm{k}}+\varepsilon_{n\bm{k}}-2\mu}{2}\right)^{\alpha+\beta}|\partial_{k^\nu}u_{n\bm{k}}\rangle\right].
  \end{aligned}
\end{equation}
We thus identify
\begin{equation}
      \mathcal{M}^{(m)}_\rho=\frac{i\epsilon_{\rho\mu\nu}}{\hbar}\int\frac{d^2k}{(2\pi)^2}\sum_{n\in\mathrm{occ}}\langle\partial_{k_\mu}u_{n\bm{k}}|\left(\frac{\hat{H}_{\bm{k}}+\varepsilon_{n\bm{k}}-2\mu}{2}\right)^{m}|\partial_{k_\nu}u_{n\bm{k}}\rangle\,,
\end{equation}
hence recovering Eq.~\eqref{eq:hier_k}.
\subsection{Partitioning the generalized magnetizations}
The partition of the generalized magnetization into sum rules arises from the mathematical relation 
\begin{equation}
    \begin{aligned}
        \left(\varepsilon_{n\bm{k}}-\mu\right)^m&=\left[\left(\frac{\varepsilon_{n\bm{k}}+\hat{H}_{\bm{k}}-2\mu}{2}\right)+\left(\frac{\varepsilon_{n\bm{k}}-\hat{H}_{\bm{k}}}{2}\right)\right]^m\\
        &=\sum_{k=0}^m\begin{pmatrix}
            m\\k
        \end{pmatrix}\left(-\frac{1}{2}\right)^k\left(\hat{H}_{\bm{k}}-\varepsilon_{n\bm{k}}\right)^k\left(\frac{\varepsilon_{n\bm{k}}+\hat{H}_{\bm{k}}-2\mu}{2}\right)^{m-k}\,,
    \end{aligned}
\end{equation}
such that
\begin{equation}
    \begin{aligned}
        \mathcal{M}^{(m)}_\rho&=\frac{i\epsilon_{\rho\mu\nu}}{\hbar}\int\frac{d^2k}{(2\pi)^2}\sum_{n\in\mathrm{occ}}\left(\varepsilon_{n\bm{k}}-\mu\right)^m\langle\partial_{k_\mu}u_{n\bm{k}}|\partial_{k_\nu}u_{n\bm{k}}\rangle\\
        &-\sum_{k=1}^m\begin{pmatrix}
            m\\k
        \end{pmatrix}\left(-\frac{1}{2}\right)^k\frac{i\epsilon_{\rho\mu\nu}}{\hbar}\int\frac{d^2k}{(2\pi)^2}\sum_{n\in\mathrm{occ}}\langle\partial_{k_\mu}u_{n\bm{k}}|\left(\hat{H}_{\bm{k}}-\varepsilon_{n\bm{k}}\right)^k\left(\frac{\varepsilon_{n\bm{k}}+\hat{H}_{\bm{k}}-2\mu}{2}\right)^{m-k}|\partial_{k_\nu}u_{n\bm{k}}\rangle\,.
    \end{aligned}
\end{equation}
We thus identify the natural decomposition
\begin{equation}
    \mathcal{M}^{(m)}_\rho=\mathcal{M}_\rho^{COM,(m)}-\sum_{k=1}^m\begin{pmatrix}
            m\\k
        \end{pmatrix}\left(-\frac{1}{2}\right)^km_\rho^{m-k,k}\,,
\end{equation}
recovering Eq.~\eqref{general_partition} from the main text. The center of mass contribution $\mathcal{M}_\rho^{COM,(m)}$ generalizes Eqs.~\eqref{orb_COM} and \eqref{Mq3:com} to higher orders,
\begin{equation}
    \begin{aligned}
        \mathcal{M}_\rho^{COM,(m)}&=\frac{i\epsilon_{\rho\mu\nu}}{\hbar}\int\frac{d^2k}{(2\pi)^2}\sum_{n\in\mathrm{occ}}\left(\varepsilon_{n\bm{k}}-\mu\right)^m\langle\partial_{k_\mu}u_{n\bm{k}}|\partial_{k_\nu}u_{n\bm{k}}\rangle\\
        &=\frac{1}{\hbar}\int\frac{d^2k}{(2\pi)^2}\sum_{n\in\mathrm{occ}}\left(\varepsilon_{n\bm{k}}-\mu\right)^m\Omega^\rho_{n\bm{k}}\,,
    \end{aligned}
\end{equation}
while the other terms $m_\rho^{m-k,k}$ with $k>0$ are generalized sum rules for the $m-k$ magnetizations. Indeed
\begin{equation}
    \begin{aligned}
    \label{eq:app_gene_sum_rule}
        m_\rho^{m-k,k}&=\frac{i\epsilon_{\rho\mu\nu}}{\hbar}\int\frac{d^2k}{(2\pi)^2}\sum_{n\in\mathrm{occ}}\langle\partial_{k_\mu}u_{n\bm{k}}|\left(\hat{H}_{\bm{k}}-\varepsilon_{n\bm{k}}\right)^k\left(\frac{\varepsilon_{n\bm{k}}+\hat{H}_{\bm{k}}-2\mu}{2}\right)^{m-k}|\partial_{k_\nu}u_{n\bm{k}}\rangle\\
        &=\int_0^{+\infty}d\omega\,(\hbar\omega)^k\frac{i\epsilon_{\rho\mu\nu}}{\hbar}\int\frac{d^2k}{(2\pi)^2}\sum_{n\in\mathrm{occ}}\sum_{m\in\mathrm{unocc}}\langle\partial_{k_\mu}u_{n\bm{k}}|u_{m\bm{k}}\rangle\langle u_{m\bm{k}}|\left(\frac{\varepsilon_{n\bm{k}}+\hat{H}_{\bm{k}}-2\mu}{2}\right)^{m-k}|\partial_{k_\nu}u_{n\bm{k}}\rangle\delta(\omega-\omega_{mn})\\
        &=\int_0^{+\infty}d\omega\,(\hbar\omega)^km_\rho^{m-k}(\omega),
    \end{aligned}
\end{equation}
where $m_\rho^k(\omega)$, analogously to Eq.~\eqref{freq_res_magnet}, defines a frequency-resolved magnetization since 
\begin{equation}
  \int_0^{+\infty}m^k_\rho(\omega)\,d\omega=m_\rho^{k,0}=\mathcal{M}_\rho^{(k)}\,. 
\end{equation}
Considering the dichroic measurement of this frequency-resolved magnetization, using generalized polarization operators [Eqs.~\eqref{gen_pol_eq_bis} and \eqref{gen_pol_eq_bis}], 
\begin{equation}
    \begin{aligned}
        m^k_{\rho}(\omega)&=\epsilon_{\rho\mu\nu}\frac{\Delta\Gamma^{\mu\nu}_{\alpha\beta}(\omega)}{V_{0\alpha}V_{0\beta}}\, , \qquad \alpha+\beta=k\\
        &=\frac{i\epsilon_{\rho\mu\nu}}{\hbar}\int\frac{d^2k}{(2\pi)^2}\sum_{n\in\mathrm{occ}}\sum_{m\in\mathrm{unocc}}\langle\partial_{k_\mu}u_{n\bm{k}}|u_{m\bm{k}}\rangle\langle u_{m\bm{k}}|\left(\frac{\varepsilon_{n\bm{k}}+\hat{H}_{\bm{k}}-2\mu}{2}\right)^{k}|\partial_{k_\nu}u_{n\bm{k}}\rangle\delta(\omega-\omega_{mn})\,,
    \end{aligned}
\end{equation}
we ultimately recover the relation displayed in Eq.~\eqref{eq_freq_res_gen}:
\begin{equation}
    \begin{aligned}
m_\rho^{a,b}&=\int_0^{+\infty}d\omega\,(\hbar\omega)^bm_\rho^{a}(\omega)
\\&=\epsilon_{\rho\mu\nu}\int_0^\infty d\omega\,\left(\hbar\omega\right)^b\frac{\Delta\Gamma^{\mu\nu}_{\alpha\beta}(\omega)}{V_{0\alpha}V_{0\beta}}, \quad \alpha+\beta=a .
    \end{aligned}
\end{equation}
\end{widetext}

\end{document}